\documentclass[structabstract]{aa}
\usepackage{graphicx}
\usepackage{longtable}
\usepackage{lscape}
\usepackage{rotating}
\usepackage{ulem}
\usepackage{txfonts}
\def\arcsec{$^{\prime\prime}$}

\begin{document}
   \title{Central Stars of Planetary Nebulae: New spectral classifications and catalogue
\thanks{Based on data collected at the Complejo Astron\'omico
El Leoncito (CASLEO), which is operated under agreement between the
Consejo Nacional de Investigaciones Cient\'{i}ficas y T\'ecnicas de
la Rep\'ublica Argentina y Universidades Nacionales de La Plata,
C\'ordoba y San Juan, Argentina.}}


   \author{W. A. Weidmann
          \inst{1}\fnmsep\thanks{Fellow of Consejo Nacional de Investigaciones
               Cient\'ificas y T\'ecnicas (CONICET), Argentina.}
    \and
          R. Gamen\inst{2}\fnmsep\thanks{Member of Carrera del Investigador CONICET,
          Argentina.}
          }

   \institute{Observatorio Astron\'omico C\'ordoba,
              Universidad Nacional de C\'ordoba, Argentina\\
              \email{walter@mail.oac.uncor.edu}
    \and
             Instituto de Astrof\'isica de La Plata, (CCT La Plata-CONICET, Universidad Nacional de La Plata)\\
             \email{rgamen@fcaglp.unlp.edu.ar}
             }



  \abstract
   {There are more than 3000 true and probable known 
   Galactic Planetary Nebulae (PNe), but 
   only for 13\% of them there is central star spectroscopic information available.} %
            {To contribute to the knowledge of central stars of planetary nebulae
            and star evolution.}
            {We undertook a spectroscopic survey of central stars of 
            PNe in low resolution and 
            compiled a large list of central stars for which
            information was dispersed in the literature.}
             {We complete a catalogue of 492 true and probable CSPN and
              we provide a preliminary spectral classification for 45 central star of PNe, 
              This made it possible to update the proportion of CSPN with atmosphere  
              poor in hydrogen with regard to the whole in at least 30\%
              and contribute with statistical information that allow to 
              infer the origin of H-poor stars.}
           {}


   \maketitle
%

\section{Introduction}
\label{intro}


Planetary Nebula is the most luminous  
transitory phase in the life of low and intermediate mass
stars (0.6M$_{\odot}$$<$M$<$8M$_{\odot}$) 
on their evolution from the Asymptotic Giant Branch (AGB) 
to their final destiny, the white dwarfs (WD). 
The PN phase begins once the central star reaches 
an effective temperature of 30000~K and 
ionises the shell of material ejected 
during its evolution in the AGB.
After about about $2\times 10^4$ years, 
it ends when the nuclear burning in a thin shell 
of the star stops, and the nebula finally disperses.

PNe were discovered more than two centuries ago, and
their number has increased every year, but there are 
still unsolved questions about them.
Some of these, and perhaps the most important ones, 
are related to aspects of the central stars
of the Planetary Nebulae (CSPN). 
The Planetary Nebulae nuclei are not located in a 
confined region of the HR diagram, moreover  
their optical spectra encompass all varieties known 
for hot stars, i.e. ranging from pure emission 
through emission-absorption mixtures and from 
near-continuous to pure strong absorption. 
The appearance of the spectrum depends upon temperature, 
luminosity, and chemical composition, or more fundamentally, 
upon core mass and state of evolution.
At the beginning of '90  M\'endez (\cite{men91}) suggested
that the majority of CSPN can be classified
in two well differentiated categories: those that 
stellar H features can be identified in their spectra 
(Hydrogen-rich) and those that do not (Hydrogen-poor).

At the present, there are about 3000 true
and probable PNe known in our Milky Way, 
listed in Acker et al. 1992, 1996 
(SECGPN\footnote{Strasbourg-ESO Catalogue of Galactic PN (SECGPN)
http://vizier.u-strasbg.fr/viz-bin/VizieR Planetary\_Nebulae V/84/cstar}), 
Parker et al. \cite{park} and Miszalski et al. \cite{mash2} 
(MASH\footnote{Macquarie/AAO/Strasbourg H$\alpha$ Planetary Galactic Catalog
http://vizier.u-strasbg.fr/vizier/MASH}), 
and Drew et al. \cite{drew} (IPHAS, INT Photometric H-Alpha Survey).
However, spectroscopic information on their central stars  
is known only in a very small fraction of objects 
(about 13\%, see Sect.~\ref{cata}).

Spectroscopy of CSPN is difficult to obtain because of their
apparent low brightness, low apparent magnitudes (60\% 
of the CSPN listed in the SECGPN are V$>15.5$) 
and the surrounding gaseous shell whose emission lines 
often mask the stellar lines. 
Besides the position of the CSPN is not always clear.

The determination of spectral types of CSPN 
should help significantly to improve our knowlege 
of their general evolutionary scheme, making 
it possible to consider CSPN as physical objects 
with individual parameters and peculiarities 
and not just as sources of ionizing radiation.

One of the first lists of CSPN was compiled by 
Aller (\cite{all48}), then by Acker et al. 1982 
(Catalogue of CSPN, Strasbourg Observatory). 
Information on CSPN can be found in the 
SECGPN and the MASH CDS-catalogues. 
Several authors added contributions, 
although targeting particular spectral types, e.g
WR+wels (Acker \& Neiner \cite{an03}), 
B[e] (Lamers et al. \cite{lam98}), 
evolved CSPN (Napiwotzki \cite{napi99}) and
PG 1159 (Werner \& Herwig \cite{wehe06}).

In order to contribute to the knowledge of the final 
stellar evolution stages we undertook a spectroscopic survey 
of CSPN and compiled a large list of CSPN. 
The motivation of the present work lies in a series of 
astronomical   concerns: the complicated puzzle of different 
types of CSPN observed (see Table~\ref{star1}), 
few stars with spectral information, 
a lack of consensus in the evolutionary sequence of the CSPN 
and the surprising bipolarity in the hydrogen abundance.

This paper is organized as follows.
The sample and observations are described in Sect.~\ref{obs};
in Sect.~\ref{resultado} we comment about the spectral 
classification; in Sect.~\ref{cata} we present the catalogue 
of CSPN and we give a brief discussion. Finally in
Sect.~\ref{con} we present our conclusions.


\section{New spectral classification }

\subsection{Observations}\label{obs}

We have observed 45 southern CSPN selected from 
SECGPN and Boumis et al.~(\cite{boumi}), the 
coordinates of CSPN were resorted from Kerber et al.~(\cite{coor}).

The observations were carried out during a three year campaign 
between 2005 November and 2008 December that 
included a total of 31 nights of observations.
For this survey we used the REOSC spectrograph attached
to the 2.15-m telescope at CASLEO, Argentina.

A 300~line mm$^{-1}$ grating was used, which 
yielded a dispersion of 3.4~\AA\, pixel$^{-1}$.
In some nights, a grating of 600~line mm$^{-1}$ 
was used (1.6~\AA\, pixel$^{-1}$).
The gratings provide a typical wavelength
range of 3500--7000~\AA\ (3875-5530~\AA\ in the 
best resolution). The slit was opened to 
3\arcsec, consistent with the seeing at the site.


\subsection{Results}
\label{resultado}

In this first work, we present a very preliminary classification 
of the observed CSPN.
We distinguish between CSPN with absorption and emission lines.
In the former group we identified basically absorption lines 
of \ion{He}{i} and \ion{He}{ii}, such CSPN were classified as OB. 
In the next group were identified emission lines, mainly of 
CIII (4650\AA \ and 5696\AA) and \ion{C}{iv} (5806\AA), 
typical of  [WC] stars. This CSPN were classified as ``emission-line''.
We have obtained some spectra whose stellar continuum 
shows a reasonable S/N but no feature, neither absorption 
nor emission is observed. In these cases, although we 
classified them as ``continuous'' type, we expect them 
to be H-rich objects (Kudritzki et al. \cite{kudri81}).
Result are shown in Table~\ref{sample}. In a forthcoming 
paper, we will perform a detailed spectroscopic analysis.


\section{The catalogue of CSPN}\label{cata}

\subsection{Content}

Taking into account that the information about CSPN spectral 
types is scattered among many publications,
we have carried out an extensive bibliographic compilation 
of the CSPN data with the goal of producing an 
updated list of those stars that have spectroscopic information.
This list includes 492 stars of true and possible PN
with spectral-type determinations, 
45 of them are from our own new data.

Note that Parker et al.~(\cite{park})
estimated that $\sim$ 30\% 
of the MASH entries have candidate CSPN,
with about half of these representing high quality 
candidates suitable for immediate follow-up.

Transition objects, as post-AGB, PPN or 
young-PN (Ej. V 348 Sgr, CRL 618, He 1-5, 
BD+33 2642, LS IV-12 111 and He 3-1475)
were not included.

The information included in the catalogue,
discriminated by true and possible PN 
(Table~\ref{ample2}, on-line only), is:

\begin{description}

\item[c1] The PN G designation, taken from SECGPN.

\item[c2] The common name of the object. 

\item[c3-4]The equatorial coordinates (J2000.0) of the 
nebula, since in most cases there is not information 
on the position of the CSPN. Though in many cases 
this is evident,  in others it is not.

\item[c5] The spectral classification of the CSPN. 
If there are more than one, they are separated by 
a semi-colon (idem in the references column).
However we use only two spectral classification if 
is it necessary, for example when the spectral 
classification are very different.
We indicated as H-rich when the authors observed 
absorption of the Balmer series. 
In some cases the authors do not give the spectral type 
of the CS, but they describe the identified lines.
We also include the CSPN with the indications done by 
Miszalski et al.~(\cite{mash2}) in the MASHII catalogue: 
Blue, [WR] or wels. Note that the Blue aspect of CSPN 
images in MASHII is not based on spectroscopic study.

\item[c6] The reference where the spectral type 
was found (t.w. means this work).

\item[c7] The reference which mentions whether the star is 
part of a binary system (nothing if none).
Although some CSPN present a late MK spectral 
type, it is accepted that the excitation source of the PN
(if star and nebulae are physically associated)
is a hitherto undetected hot star (Lutz~\cite{latemk}). 
In those cases, we indicated bc-CSPN: binariy for the cool CSPN.

\end{description}

\begin{table}
\caption[]{Summary of the spectral types of CSPN compiled 
in our catalogue, grouped by their atmospheric hydrogen abundance.
Here, we discarded 9 objects without specific spectral type.}
\label{star1}
\centering
\setlength{\tabcolsep}{0.5mm}
\begin{tabular}{lc|lc|lc}
\hline\hline\noalign{\smallskip}
\multicolumn{2}{c}{H-rich} \vline   &     &     &  \multicolumn{2}{c}{H-poor}  \\
 Sp.Type             & sample       & Sp. Type  & sample & Sp. Type  &  sample\\
\noalign{\smallskip}\hline\noalign{\smallskip}
O3-9+B$_{early}$ & 64 &  sdB         & 1  & [WC4-11]    & 57 \\
Of            & 20 & hybrid          & 3  & [WO1-4]     & 33 \\
later that B5 & 38 & symbiotic star? & 7  & [WR]        & 11 \\
B[e]          & 6  & Blue            & 50 & [WN]        & 5 \\
DA+WD         & 12 & emission-line   & 25 & PG 1159     & 15 \\
DAO           & 14 &                 &    & [WC]-PG1159 & 2 \\
sdO           & 3  &                 &    & O(He)       & 3 \\
hgO(H)        & 16 &                 &    & O(c)+Of(c)  & 2 \\
cont.         & 16 &                 &    & H-poor      & 1 \\
H-rich        & 3  &                 &    & DO          & 4 \\
              &    &                 &    & wels        & 72 \\
\hline
total         & 192 & total          & 86 & total      & 205 \\
  \hline
\end{tabular}
\end{table}

The catalogue of Acker et al.~(\cite{ack92}) and AN03
provided spectroscopic information of 240 CSPN, 
with this new collected list the number of CSPN with 
spectral classification has been doubled. 
We hope that this new list will be useful for future works.


\subsection{Discussion}
\label{DR}

The larger sample of CSPN with spectral types allows us to 
discuss about the dichotomy between H rich and poor stars.

We have grouped the H-rich and H-poor CSPN in 
Table~\ref{star1}\footnote{Although we have included 
the wels into H-poor group (because of we found 
evidence in this sense), we prefer to be cautious and 
separate in three groups: H-rich, H-poor and wels 
in the following discussion.}, It is clear that the 
former group is more numerous than the H-poor one, 
the ratio is 1.4. In an earlier study M\'endez (\cite{men91}) 
reported a ratio of 3.
It is evident that stars with strong emission lines are 
easier to detect than those with absorption lines, thus 
favouring the detection of H-poor stars. But, is this 
effect enough to explain the ratio of stars observed 
between both groups?

We obtained that above 30\% of the whole CSPNe population
appears to be hydrogen deficient (without counting the ``Blue'' stars).
It is difficult to get a theoretical prediction of 
such ratio of stellar types because the mechanism 
to generate H-poor CSPNe is not well known.
The more accepted hypothesis to explain the lack 
of hydrogen in the atmospheres of CSPN is the born-again 
phenomenon (Iben et al.~\cite{ibe}).
In this framework, it is estimated that roughly a 15\% 
(Lawlor \& MacDonald~\cite{lawM}) of post-AGB 
stars suffer a born-again event.
Bl\"ocker et al. (\cite{bow}), based on their 
improved born-again models (thermal pulses plus overshooting), 
obtained that 20\% to 25\% of stars can be expected to 
become H-poor. These theoretical values are quite lower 
than our observational one. 
According to this catalogue, it is difficult to 
imagine that a selection effect could be as 
efficient as to produce such high fraction of 
H-poor star, so perhaps the born-again phenomenon is not 
the unique mechanism to obtain an atmosphere free of hydrogen. 
We recall other ways to get H-poor CSPN, as the 
binary channel (Tylenda \& G\'orny, \cite{tyl}) or the 
continuous stripping of the outer H-rich layers by 
intense stellar winds (G\'orny \& Tylenda, \cite{gor00}).

Only 71 close binary CSPN are known nowadays 
(de Marco~\cite{d2009}, Miszalski et al.~\cite{MAm2009}
and~\cite{MC2010}), in addition it is interesting 
to note that almost all of them present an H-rich spectra.
Recently was discovered the first [WR] star in a close 
binary system (Hajduk et al.~\cite{HZ2010}).
Note that nearly 14\% of the compiled CSPN are 
probable binary systems, in good agreement with the 
10-15 value obtained by Bond et al.~(\cite{bcfg}).

\begin{figure*}
   \centering
   \includegraphics[width=7.5cm]{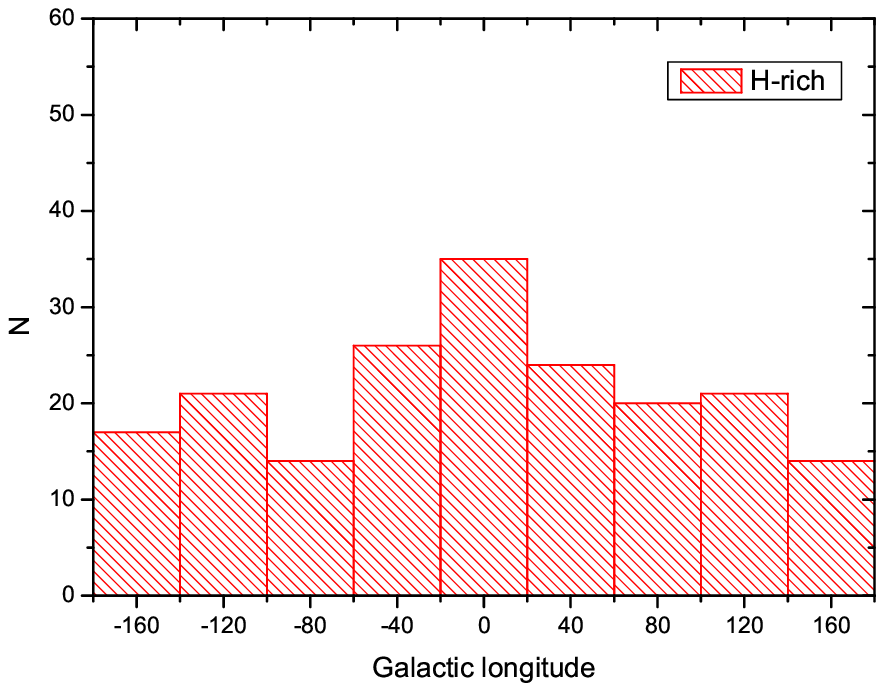}
   \includegraphics[width=7.5cm]{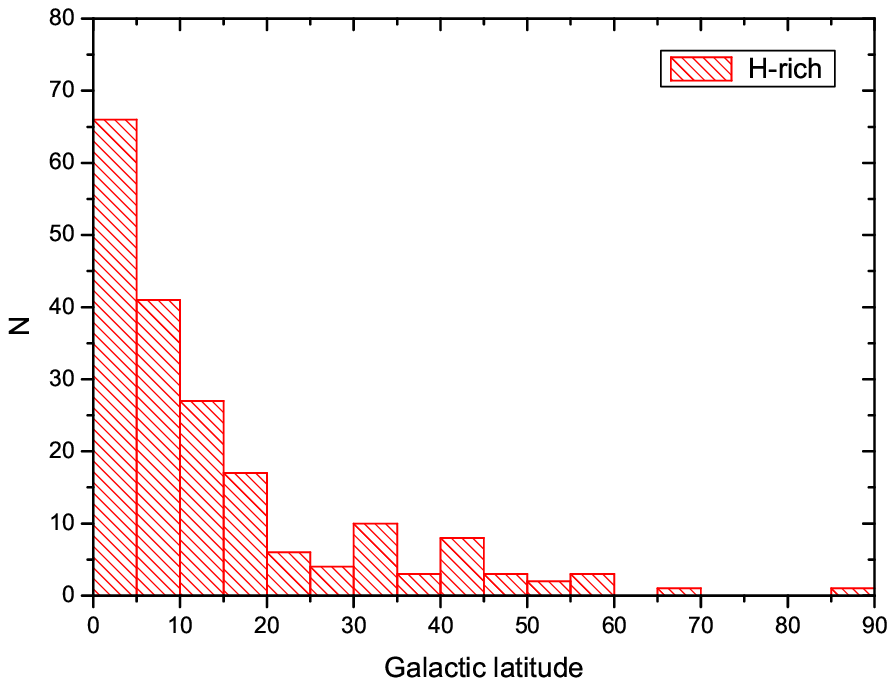}

   \includegraphics[width=7.5cm]{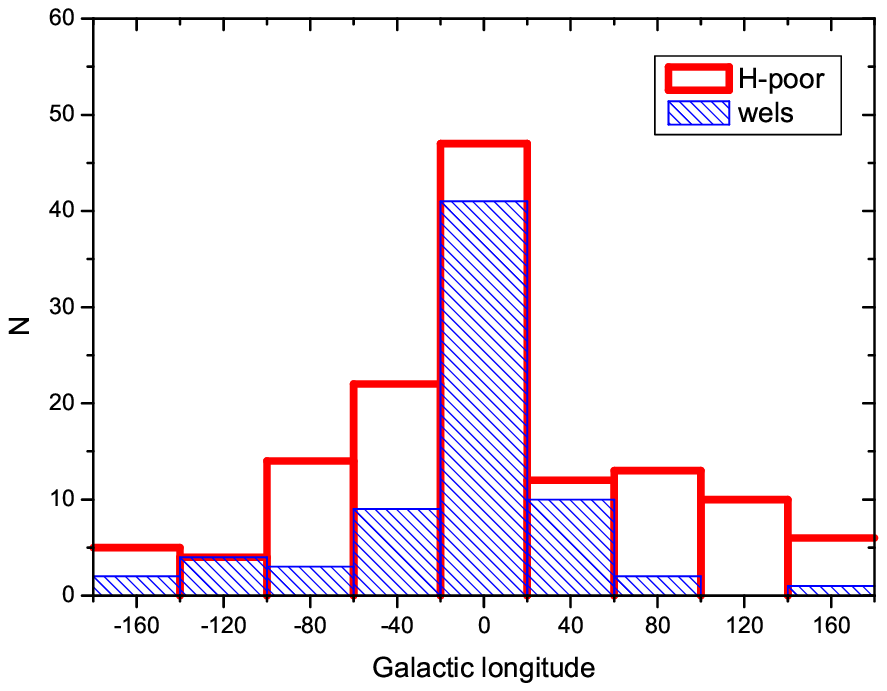}
   \includegraphics[width=7.5cm]{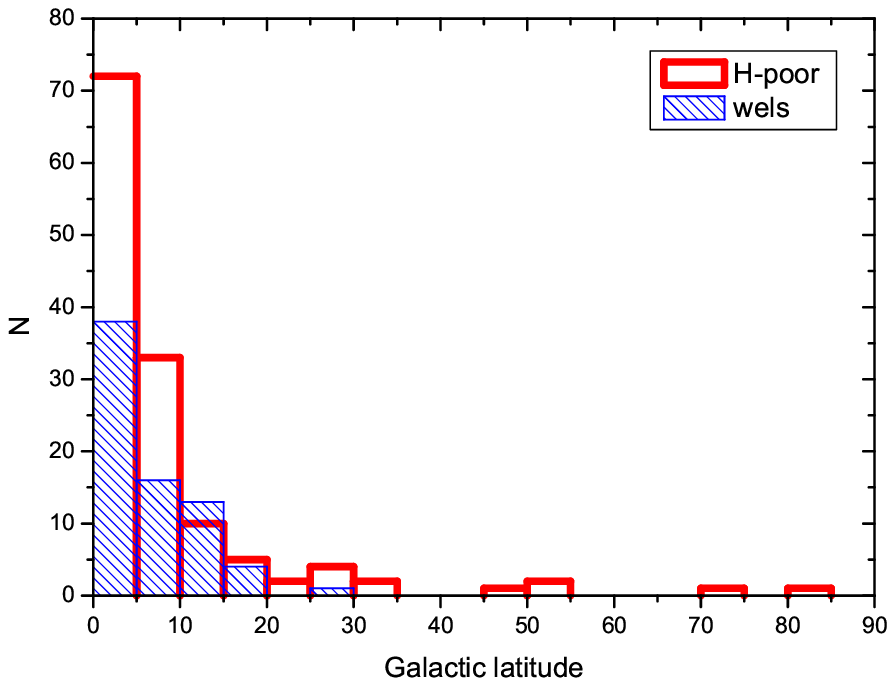}

   \caption[]{Distribution in Galactic longitude and latitude
             of CSPN (of true and possible PN) that belong to H-rich, 
             H-poor and wels stars groups.
             Note that H-poor PN are more concentrated towards 
             the Galactic center than H-rich ones.
             It is also noticeable the similarity between wels and H-poor distributions.}
     \label{l-stat}
   \end{figure*}

We analysed the distribution in galactic coordinates of the 
CSPN sample that belongs to H-rich and H-poor group. 
From Fig.~\ref{l-stat}, it is evident that
a strong concentration of H-poor and wels stars toward the 
Galactic center exists. This effect has been observed by 
G\'orny et al. (\cite{gor}) and attributed to a possible 
selection effect. However, this effect could be 
interpreted as an influence of the metallicity 
in the mechanism that unleashes the total hydrogen 
loss from the stellar atmosphere of those objects.

On the other hand, the average height above the Galactic plane,
of H-rich, H-poor and wels stars was found to be 
13.9$^\circ \pm$ 15.2, 9.0$^\circ \pm$ 12.6 and 6.7$^\circ \pm$ 5.3
respectively. As these errors are too high, we
undertaken a Kolmogorov-Smirnov (KS) statistical analysis.
The significance of trends of KS test is assessed on 
the basis of differences, D, between their cumulative 
number distributions. This is used to define a probability 
coeficient P, such that low values of P imply 
significant levels of difference.
The result of KS test is show in Table~\ref{ks},
it is clear that the distribution of galactic latitudes 
of H-rich and H-poor stars are very different. In
addition, the sample of wels stars have, apparently, 
more similarities with the H-poor star than the other 
group, supporting the hypothesis that wels belong 
to the H-poor group and increasing even more the 
ratio of H-poor over the whole CSPN.

\begin{table}
\caption{Summary of result of KS test applied to the sample of
         galactic latitude. Where D indicate the differences 
         between the cumulative number distributions
         and P the probability that the compared samples are equal.}
\label{ks}
\centering
\begin{tabular}{l|c|r}
\hline\hline\noalign{\smallskip}
 \ \ \ \ \ \ compared    &     &    \\
\ \ \ \ \ \ \ \ groups   &  D  & P \ \ \ \\
\noalign{\smallskip}\hline\noalign{\smallskip}
H-rich vs. H-poor   &   0.26    &  $<$0.1\%   \\
H-rich vs. wels     &   0.25    &  0.3\%   \\
H-poor vs. wels     &   0.11    &  64.1\%   \\
  \hline
\end{tabular}
\end{table}


\section{Conclusions}
\label{con}

We have carried out a spectroscopic survey of PN, 
where we have performed a very preliminary
determination of the spectral types of 45 of its 
central stars, all of them previously unclassified.
In addition, we have performed an extensive bibliographic 
compilation of CSPN with determined spectral types.
Thus, we present the list of 492 CSPN which are classified 
(together with their respective references), and also 
included a tag indicating those binary systems or candidates.  
We hope this list will be useful for future investigations.

From our catalogue we grouped CSPN whose atmospheres are 
hydrogen rich or poor; conservatively we ruled out the wels
(nevertheless we found evidence supporting the hypothesis 
that wels belong to H-poor group). 
We found that the ratio of stars in both groups is 
lower than previous estimations.
According to our recent statistical analysis, we found 
that PN with H-poor central star are more concentrated
toward the galactic center and galactic plane than 
the H-rich group. It suggests that H-poor stars could 
have a more massive progenitor and besides, the 
metallicity could play an important role in the mechanism 
responsible for originating  hydrogen-free atmospheres.
In addition we found that the frequency of 
occurrence of known close binaries among CSPNe is $\sim$14\%.


\begin{acknowledgements}
We would like to thank our anonymous referee
whose critical remarks helped us to substantially
improve this paper.
The CCD and data acquisition system at CASLEO has been
financed by R. M. Rich trough U. S. NSF grant AST-90-15827. 
This work has been partially supported by Concejo de Investigaciones
Cientif\'{i}cas y T\'ecnicas de la Rep\'ublica Argentina (CONICET).
This research has made use of the SIMBAD database, 
operated at CDS, Strasbourg, France.
The authors wish to thank Drs. Guillermo Bosch and Roberto H. M\'endez for
comments that helped to improve the paper. 
\end{acknowledgements}


\longtab{3}{
\begin{longtable}{lccccc}
\caption[]{Spectral types from our observations. The PN are denoted by 
           their common name and by their PN G designation.
           Fifth column list the preliminary spectral type 
           that we have adopted for each CSPN.
           In last column is indicate the exposure time and grating used
           ( $300$ or $600$~line mm$^{-1}$).
}\label{sample} \\
\hline\hline
name &  PN G & AR(2000) & DEC(2000) & Sp. Type   &    E.T.[s] (grating)      \\
\hline
\endfirsthead
\caption{continued.}\\
\hline\hline
name & PN G  & AR(2000) & DEC(2000)  & Sp. Type  &     E.T.[s] (g.g.)    \\
\hline
\endhead
\hline
\endfoot
H 1-62  & 000.0-06.8 & 18 13 17.9 & -32 19 43.0 & emission-line & 3600 (300)      \\
PC 12   & 000.1+17.2 & 16 43 49.3 & -18 56 33.0 &  OB           & 2$\times$1200 (300)  \\
IC 4634 & 000.3+12.2 & 17 01 33.5 & -21 49 33.1 & emission-line & 3$\times$1000 (300) \\
H 1-63  & 002.2-06.3 & 18 16 18.5 & -30 07 35.8 & OB?          & 3600 (300)       \\
M 1-38  & 002.4-03.7 & 18 06 05.8 & -28 40 34.3 & cont.        &  3600 (300)      \\
M 1-53  & 015.4-04.5 & 18 35 48.2 & -17 36 08.4 & emission-line? &  3600 (300)    \\
Sa 1-8  & 020.7-05.9 & 18 50 44.2 & -13 31 02.4 &  OB          &  3600 (300)      \\
IRAS 19021+0209 & 036.4-01.9 & 19 04 38.5 & 02 14 23.0 & cont. & 3600 (300)       \\
M 1-6   & 211.2-03.5 & 06 35 44.6 & -00 05 41.1 & emission-line &  3600 (300)     \\ 
SaSt 2-3 & 232.0+05.7 & 07 48 03.5 & -14 07 42.6 & OB          &  3600 (300)      \\ 
M 1-11  & 232.8-04.7 & 07 11 16.6 & -19 51 03.0 & emission-line &  3600 (300)     \\  
M 1-14  &  234.9-01.4 & 07 27 56.5 & -20 13 23.4 & OB          &  2$\times$3600 (300)  \\  
M 1-12  &  235.3-03.9 & 07 19 21.4 & -21 43 55.3 & emission-line &  3600 (300)    \\
Y-C 2-5 & 240.3+07.0 & 08 10 41.7 & -20 31 32.9  & emission-line &  3600 (300)    \\
KLSS 1-9 & 240.8-19.6 & 06 24 36.4 & -33 04 49.0 & OB          &   3600 (300)     \\
M 3-4   & 241.0+02.3 & 07 55 11.2 & -23 37 45.6 & cont.        &   3600 (300)     \\
M 3-1   & 242.6-11.6 & 07 02 49.6 & -31 35 41.3 & cont.        &   3600 (300)     \\
M 4-2   & 248.8-08.5 & 07 28 55.2 & -35 45 15.4 & emission-line &   3600 (300)    \\
Ns 238  & 254.6+00.2 & 08 20 56.7 & -36 13 46.7 & OB           &  2$\times$3600 (300)   \\
PB 2    & 263.0-05.5 & 08 20 39.8 & -46 20 13.2 & emission-line?  &  2$\times$1200 (300)  \\
PB 4    & 275.0-04.1 & 09 15 07.6 & -54 52 38.5 & emission-line?  &  3600 (300)   \\  
IC 2501 & 281.0-05.6 & 09 38 47.5 & -60 05 27.9 & emission-line   &  2$\times$3600 (300)   \\
IC 2553 & 285.4-05.3 & 10 09 21.7 & -62 36 40.9 & emission-line   &  4$\times$300 (300)   \\ 
He 2-47 & 285.6-02.7 & 10 23 09.0 & -60 32 34.3 & emission-line   &  2$\times$2700 (300)  \\
IC 2621 & 291.6-04.8 & 11 00 19.5 & -65 14 54.2 & emission-line   &  2$\times$3600 (300)    \\ 
Lo 6    & 294.1+14.4 & 12 00 43.5 & -47 33 12.0 & cont.           &  3600 (300)   \\
Th 2-A  & 306.4-00.6  & 13 22 34.8 & -63 20 55.2 & emission-line  &  3600 (300)\footnote{GEMINI observation, see Weidmann et al. (\cite{ww})}    \\
He 2-97 & 307.2-09.0 & 13 45 24.0 & -71 28 48.8 & emission-line   &   3600 (300)  \\  
He 2-105 & 308.6-12.2 & 14 15 25.7 & -74 12 49.8 &  OB            &   3600 (300)  \\  
NGC 5307 & 312.3+10.5 & 13 51 03.3 & -51 12 15.9 & emission-line  &   3600 (300)  \\   
He 2-107 & 312.6-01.8 & 14 18 42.5 & -63 07 10.7 & emission-line  &   2$\times$3600 (300)    \\
He 2-434 & 320.3-28.8 &  19 33 50.7 & -74 32 58.7 & OB            &   3600 (300)   \\ 
NGC 5979 & 322.5-05.2 & 15 47 40.6 & -61 13 02.7 & emission-line  &   2$\times$1500 (300)    \\
He 2-128 & 325.8+04.5 & 15 25 07.9 & -51 19 40.9 & emission-line? &   3600 (600)  \\  
WRAY 17-75 & 329.5-02.2 & 16 12 34.4 & -54 23 35.3 & OB           &   3600 (300)  \\  
He 2-187 & 337.5-05.1 & 17 01 37.4 & -50 22 56.6 & OB             &   3600 (300)  \\
NGC 6026 & 341.6+13.7 & 16 01 20.8 & -34 32 38.0 & OB             &   3600 (300)  \\
PC 17    & 343.5-07.8 & 17 35 41.1 & -46 59 51.3 & emission-line  &   3600 (600)       \\
Cn 1-3   & 345.0-04.9 & 17 26 11.8 & -44 11 29.1 & emission-line? &   4$\times$700 (300)    \\
IC 4663  & 346.2-08.2 & 17 45 28.5 & -44 54 11.5 & emission-line? &   4$\times$700 (600)  \\
IC 4699  & 348.0-13.8 & 18 18 31.2 & -45 59 03.2 & emission-line  &   3600 (600)  \\ 
NGC 6337 & 349.3-01.1 & 17 22 16.0 & -38 28 57.6 & emission line  &   3600 (300)  \\  
Fg 3     & 352.9-07.5 & 18 00 11.9 & -38 49 51.7 & cont.          &   3600 (300)  \\
H 1-35   & 355.7-03.5 & 17 49 13.9 & -34 22 53.3 & emission-line? &   2$\times$1700 (300) \\
Te 2022  & 358.8-00.0 & 17 42 42.4 & -29 51 35.4 & OB             &   3600 (300)  \\
\end{longtable}
}



\begin{thebibliography}{}


\bibitem[1966]{ab66}  Abell, G. O. 1966, ApJ, 144, 259

\bibitem[1982]{ack82} Acker, A., Gleizes, F., Chopinet, M., Marcout, J., Ochsenbein, F., Roques, J. M. 1982, CDSSP, 3

\bibitem[1992]{ack92} Acker, A., Marcout, J., Ochsenbein, F., Stenholm, B. \& Tylenda, R. 1992
                      Strasbourg-ESO catalogue of galactic planetary nebulae.
                      Garching: European Southern Observatory, 1992-1996.

\bibitem[2003]{an03}   Acker, A. \& Neiner, C. 2003, A\&A, 403, 659

\bibitem[1948]{all48}  Aller, L. H., 1948, ApJ, 108, 462

\bibitem[1985]{AK1985} Aller, L. H. \& Keyes, C. D.  1985, PASP, 97, 1142

\bibitem[1987]{AK1987} Aller, L. H. \& Keyes, C. D.  1987, ApJS, 65, 405

\bibitem[1996]{all96}  Aller, L. H., Hyung, S., Feibelman, W. A. 1996, PASP, 108, 488

\bibitem[2000]{bel00} Belczy$\acute{n}$ski, K., Mikolajewska, J., Munari, U., 
Ivison, R. J., Friedjung, M. 2000, A\&AS, 146, 407

\bibitem[2003]{BC2003} Benetti, S., Cappellaro, E., Ragazzoni, R., Sabbadin, F., Turatto, M. 2003, A\&A, 400, 161

\bibitem[1993]{BD1993}  Bianchi, L. \& Defrancesco, G.  1993, IAUS, 155, 85

\bibitem[2008]{BC2008} Bilikova, J., Chu, Y.-H., Su, K., Grundell, R., Rauch, T., Volk, K. 2008, in 16th European White Dwarf Workshop, ASP Conference Series, in press

\bibitem[2001]{bow} Bl\"ocker, T., Osterbart, R., Weigelt, G., Balega, Y., 
Men'shchikov, A. 2001, ASSL, 265, 241

\bibitem[2008]{B2008}  Bohigas, J.  2008, ApJ, 674, 954

\bibitem[1987]{BG1987} Bond, H. E. \& Grauer, A. D. 1987, IAU Colloq. 95: Second Conference on Faint Blue Stars, ed. A. G. D. Philip, D. S. Hayes, J. W. Liebert, 221–228

\bibitem[1989]{bond}   Bond, H. E., Ciardullo, R., Meakes, M. 1989, BAAS, 21, 789

\bibitem[1989]{bcfg} Bond, H. E., Ciardullo, R., Fleming, T. A., Grauer, A. D. 1989, IAUS, 131, 310

\bibitem[1990]{boli}   Bond, H. E. \& Livio, M. 1990, ApJ, 355, 568

\bibitem[1993]{BM1993} Bond, H. E., Meakes, M. G., Liebert, J. W., Renzini, A. 1993, IAUS  155, 499

\bibitem[1999]{BC1999} Bond, H. E. \& Ciardullo, R. 1999, PASP, 111, 217

\bibitem[2002]{BP2002} Bond, H. E. \& Pollacco, D. L. 2002, AP\&SS, 279, 31

\bibitem[2002]{BO2002} Bond, H. E., O'Brien, M. S., Sion, E. M., Mullan, D. J., Exter, K., Pollacco, D. L., Webbink, R. F. 2002, ASPC, 279, 239

\bibitem[2003]{BP2003} Bond, H. E., Pollacco, D. L., Webbink, R. F.  2003, AJ, 125, 260

\bibitem[2003]{boumi} Boumis, P., Paleologou, E. V., Mavromatakis, F.,
                      Papamastorakis, J. 2003, MNRAS, 339, 735

\bibitem[1971]{chB} Brocklehurst, M. 1971, MNRAS, 153, 471

\bibitem[1985]{CP1985}  Cerruti-Sola, M., Perinotto, M.  1985, ApJ, 291, 237

\bibitem[1980]{C1980}  Chromey, F. R.  1980, AJ, 85, 853

\bibitem[2009]{CG2009} Chu, Y., Gruendl, R. A., Guerrero, M. A., Su, K. Y. L., Bilikova, J., Cohen, M., Parker, Q. A., Volk, K., Caulet, A., Chen, W., and 2 coauthors  2009, AJ, 138, 691

\bibitem[1999]{ciar} Ciardullo, R., Bond, H. E., Sipior, M. S., Fullton, L. K.,
                     Zhang, C.-Y., Schaefer, K. G. 1999, AJ, 118, 488

\bibitem[1987]{CJ1987} Cohen, M. \& Jones, B. F.  1987, ApJ, 321, 151L

\bibitem[1995]{C1995}  Corradi, R. L. M.  1995, MNRAS, 276, 521

\bibitem[2003]{dS2003}  de Marco, O., Sandquist, E. L., Mac Low, M., M., Herwig, F., Taam, R. E.  2003, RMxAC, 18, 84

\bibitem[2006]{d2006}  de Marco, O.  2006, IAUS, 234, 111

\bibitem[2009]{d2009}  de Marco, O.  2009, PASP, 121, 316

\bibitem[1999]{D1999}  Dreizler, S.  1999, RvMA, 12, 255

\bibitem[2005]{drew} Drew, J.E., Greimel, R., Irwin, M.J., et al., 2005, MNRAS, 362, 753

\bibitem[1983]{D1983}  Drilling, J. S.  1983, ApJ, 270, L13

\bibitem[1985]{D1985}  Drilling, J. S. 1985, ApJ, 294, L107

\bibitem[1996]{DB1996} Duerbeck, H. W. \& Benetti, S.  1996, ApJ, 468, L111

\bibitem[2005]{EP2005} Exter, K. M., Pollacco, D. L., Maxted, P. F. L., Napiwotzki, R., Bell, S. A. 2005, MNRAS, 359, 315

\bibitem[1983]{FK1983} Feibelman, W. A. \& Kaler, J. B.  1983, ApJ, 269, 592

\bibitem[1994]{feibe} Feibelman, W. A. 1994, PASP, 106, 56

\bibitem[1999]{F1999} Feibelman, W. A. 1999, PASP, 111, 719

\bibitem[1981]{femcs} Ferguson, D. H., McGraw, J. T., Spinrad, H., Liebert, J., \& Green, R. F. 1981, ApJ, 251, 205

\bibitem[2006]{FP2006} Frew, D. J., Parker, Q. A., Russeil, D.  2006, MNRAS, 372, 1081

\bibitem[2010]{FS2010} Frew, D. J., Stanger, J., Fitzgerald, M., Parker, Q., Danaia, L., McKinnon, D., Guerrero, M. A., Hedberg, J., Hollow, R., An, Y., and 9 coauthors 2010, PASA, in press

\bibitem[2001]{GP2001} Gauba, G., Parthasarathy, M., Nakada, Y., Fujii, T.  2001, A\&A, 373, 572

\bibitem[2003]{GZ2003} Gesicki, K. \& Zijlstra, A. A.  2003, MNRAS, 338, 347

\bibitem[2006]{GZ2006} Gesicki, K., Zijlstra, A. A., Acker, A., G\'orny, S. K., Gozdziewski, K., Walsh, J. R.  2006, A\&A, 451, 925

\bibitem[2000]{gor00}  G\'orny, S. K. \& Tylenda, R. 2000, A\&A, 362, 1008

\bibitem[2003]{GS2003} G\'orny, S. K., Si\'odmiak, N.  2003, IAUS, 209, 43

\bibitem[2004]{gor} G\'orny, S. K., Stasinska, G., Escudero, A. V., Costa, R. D. D. 2004, 
                    A\&A, 427, 231

\bibitem[2009]{GC2009}  G\'orny, S. K., Chiappini, C., Stasinska, G., Cuisinier, F.  2009, A\&A, 500, 1089

\bibitem[1983]{GB1983} Grauer, A. D. \& Bond, H. E. 1983, ApJ, 271, 259

\bibitem[1987]{GB1987} Grauer, A. D., Bond, H. E., Ciardullo, R., Fleming, T. A. 1987, BAAS, 19, 643

\bibitem[2010]{HZ2010} Hajduk, M., Zijlstra, A., Gesicki, K. 2010, MNRAS, 406, 626

\bibitem[1992]{hamu}  Hamuy, M., Walker, A. R., Suntzeff, N. B., Gigoux, P.,
                     Heathcote, S. R., Phillips, M. M. 1992, PASP, 104, 533

\bibitem[2003]{H2003} Handler, G.  2003, IAUS, 209, 237

\bibitem[1993]{HP1993}  Harrington, J. P. \& Paltoglou, G.  1993, ApJ, 411, L103

\bibitem[1984]{HD1984}  Heber, U. \& Drilling, J. S.  1984, MitAG, 62, 252

\bibitem[2004]{HI2004}  Hewett, P. \& Irwin, M.  2004, INGN, 8, 6

\bibitem[2006]{HB2006}  Hillwig, T. C., Bond, H. E., Afsar, M. 2006, IAUS, 234, 421

\bibitem[2006]{HI2006} Hsia, C. H., Ip, W. H., Li, J. Z. 2006, AJ, 131, 3040

\bibitem[2007]{HP2007}  Hultzsch, P. J. N., Puls, J., M\'endez, R. H., Pauldrach, A. W. A., Kudritzki, R., P., Hoffmann, T. L., McCarthy, J. K.  2007, A\&A, 467, 1253

\bibitem[1999]{hyu} Hyung, S., Aller, L. H., Feibelman, W. A. 1999, ApJ, 525, 294

\bibitem[1983]{ibe} Iben, I., Jr., Kaler, J. B., Truran, J. W., Renzini, A. 1983, ApJ, 264, 605

\bibitem[1969]{JE1969}  Jones, D. H. P., Evans, D. S., Catchpole, R. M.  1969, Obs., 89, 18

\bibitem[1994]{kerf94} Kerber, F., Lercher, G., Sauer, W., Seeberger, R., 
                       Weinberger, R. 1994, AGAb, 10, 172

\bibitem[2003]{coor} Kerber, F., Mignani, R. P., Guglielmetti, F., 
                     Wicenec, A. 2003, A\&A, 408, 1029

\bibitem[1994]{KB1994}  Kingsburgh, R. L. \& Barlow, M. J.  1994, MNRAS, 271, 257

\bibitem[1994]{K1994}  Kondrateva, L. N.  1994, AstL, 20, 644

\bibitem[2005]{KB2005}  Kraus, M., Borges Fernandes, M., de Ara\'ujo, F. X., Lamers, H. J. G. L. M. 2005, A\&A, 441, 289

\bibitem[1981]{kudri81} Kudritzki, R. P., Simon, K. P., M\'endez, R. H. 1981, Msngr, 26, 7

\bibitem[1989]{KC1989}  Kwitter, K. B., Congdon, C. W., Pasachoff, J.�M., Massey, P.  1989, AJ, 97, 1423

\bibitem[1998]{lam98} Lamers, H. J. G., Zickgraf, F., de Winter, D., Houziaux, L., 
Zorec, J., 1998, A\&A, 340, 117

\bibitem[2001]{LR1983} Law, W. Y. \& Ritter, H. 1983, A\&A, 123, 33

\bibitem[2001]{lawM} Lawlor, T. M. \& MacDonald, J.  2001, ASPC, 226, 20

\bibitem[2007]{LS2007}  Lee, T-H, Stanghellini, L., Ferrario, L., Wickramasinghe, D.  2007, AJ, 133, 987

\bibitem[2000]{LS2000}  Liu, X. W., Storey, P. J., Barlow, M. J., Danziger, I. J., Cohen, M., Bryce, M.  2000, MNRAS, 312, 585

\bibitem[1977]{latemk} Lutz, J. H. 1977, A\&A, 60, 93

\bibitem[1987]{LK1987} Lutz, J. H. \& Kaler, J. B. 1987, BAAS, 19, 1090

\bibitem[2006]{MC2006} Mampaso, A., Corradi, R. L. M., Viironen, K., Leisy, P., Greimel, R., Drew, J. E., Barlow, M. J., Frew, D. J., Irwin, J., Morris, R. A. H., and 4 coauthors  2006, A\&A, 458, 203

\bibitem[1981]{MD1981} Margon, B., Downes, R. A., Katz, J. I. 1981, Natur, 293, 200

\bibitem[1988]{MK1988}  M\'endez, R. H., Kudritzki, R. P., Herrero, A., Husfeld, D., Groth, H. G.  1988, A\&A, 190, 113

\bibitem[1988]{MK1988} M\'endez, R. H., Kudritzki, R. P., Groth, H. G., Husfeld, D., Herrero, A.  1988, A\&A, 197, L25

\bibitem[1977]{MN1977} M\'endez, R. H. \& Niemela, V. S. 1977, MNRAS, 178, 409

\bibitem[1981]{MN1981} M\'endez, R. H. \& Niemela, V. S. 1981, ApJ, 250, 240

\bibitem[1982]{MN1982} M\'endez, R. H. \& Niemela, V. S.  1982, IAUS, 99, 457

\bibitem[1991]{men91}  M\'endez, R. H. 1991, IAUS, 145, 375

\bibitem[1997]{MV1997} Miranda, L. F., Vazquez, R., Torrelles, J. M., Eiroa, C., Lopez, J. A.  1997, MNRAS, 288, 777

\bibitem[2008]{mash2} Miszalski, B., Parker, Q. A., Acker, A., Birkby, J. L., Frew, D. J., Kovacevic, A. 2008, MNRAS, 384, 525

\bibitem[2009]{MAm2009} Miszalski, B., Acker, A., Moffat, A. F. J., Parker, Q. A., Udalski, A. 2009, A\&A, 496, 813

\bibitem[2009]{MA2009} Miszalski, B., Acker, A., Parker, Q. A., Moffat, A. F. J.  2009, A\&A, 505, 249

\bibitem[2010]{MC2010} Miszalski, B., Corradi, R. L. M., Jones, D., Santander-Garc\'ia, M.,  
Rodr\'iguez-Gil, P., Rubio-D\'iez, M. M., 2010, in press

\bibitem[2010]{M2010} Miszalski, B., 2010, Asymmetric Planetary Nebulae V

\bibitem[2009]{MO2007} Mitchell, D. L., O'Brien, T. J., Pollacco, D., Bryce, M.  2007, IAUS, 240, 429

\bibitem[2007]{MP2007}  Mitchell, D. L., Pollacco, D., O'Brien, T. J., Bryce, M., L\'opez, J. A., Meaburn, J., Vaytet, N. M. H.  2007, MNRAS, 374, 1404

\bibitem[2001]{MP2001}  Morgan, D. H., Parker, Q. A., Russeil, D.  2001, MNRAS, 322, 877

\bibitem[1999]{napi99} Napiwotzki, R. 1999 A\&A, 350, 101

\bibitem[2005]{nt2005} Napiwotzki, R., Tovmassian, G., Richer, M. G., Stasinska, G., Pe\~na, M., Drechsel, H., Dreizler, S., Rauch, T. 2005, AIPC, 804, 173

\bibitem[2003]{PM2003} Parker, Q. A., Morgan, D. H.  2003, MNRAS, 341, 961

\bibitem[2006]{park} Parker, Q. A., Acker, A., Frew, D. J., Hartley, M.,
                     Peyaud, A. E. J., Ochsenbein, F., Phillipps, S.,
                     Russeil, D., Beaulieu, S. F., Cohen, M.,
                     and 6 coauthors 2006, MNRAS, 373, 79

\bibitem[1992]{PT1992} Pena, M., Torres-Peimbert, S., Ruiz, M. T.  1992, A\&A, 265, 757

\bibitem[1997]{PR1997} Pena, M., Ruiz, M. T., Bergeron, P., Torres-Peimbert, S., Heathcote, S.  1997, A\&A, 317, 911

\bibitem[1995]{pe95} Pe\~na, M., Peimbert, M., Torres-Peimbert, S., Ruiz, M. T., 
                     Maza, J. 1995, ApJ, 441, 343

\bibitem[2002]{PM2002}  Pe\~na, M. \& Medina, S.  2002, RMxAA, 38, 23

\bibitem[2004]{P2004}  Pereira, C. B.  2004, A\&A, 413, 1009

\bibitem[2008]{PM2008} Pereira, C. B., Miranda, L. F., Smith, V. V., Cunha, K.  2008, A\&A, 477, 535  

\bibitem[2004]{PF2004}  Pierce, M. J., Frew, D. J., Parker, Q. A., K\"oppen, J.  2004, PASA, 21, 334  

\bibitem[1983]{P1983}  Pottasch, S. R.  1983, ASSL, 107  

\bibitem[1996]{P1996}  Pottasch, S. R.  1996, A\&A, 307  561

\bibitem[1999]{RK1999}  Rauch, T., K\"oppen, J., Napiwotzki, R., Werner, K.  1999, A\&A, 347, 169  

\bibitem[2002]{RH2002}  Rauch, T., Heber, U., Werner, K.  2002, A\&A, 381, 1007  

\bibitem[2001]{RC2001}  Rodr\'iguez, M., Corradi, R. L. M., Mampaso, A.  2001, A\&A, 377, 1042  

\bibitem[1987]{SF1987}  Sabbadin, F., Falomo, R., Ortolani, S.  1987, A\&AS, 67, 541  

\bibitem[2010]{SG2010}  Santander-Garc\'ia, M., 2010, Asymmetric Planetary Nebulae V

\bibitem[1997]{SW1997}  Saurer, W., Werner, K., Weinberger, R.  1997, A\&A, 328, 598  

\bibitem[1979]{extlaw}  Seaton, M. J. 1979, MNRAS, 187, 785

\bibitem[2004]{SL2004}  Shen, Z., X., Liu, X., W., Danziger, I. J.  2004, A\&A, 422, 563  

\bibitem[2007]{SB2007}  Smith, N., Bally, J., Walawender, J.  2007, AJ, 134, 846  

\bibitem[1997]{SZ1997}  Soker, N. \& Zucker, D. B.  1997, MNRAS, 289, 665  

\bibitem[1994]{SK1994}  Stanghellini, L., Kaler, J. B., Shaw, R. A.  1994, A\&A, 291, 604  

\bibitem[1987]{TS1987}  Tamura, S. \& Shaw, R. A.  1987, PASP, 99, 1264  

\bibitem[2004]{TN2004}  Tovmassian, G. H., Napiwotzki, R., Richer, M. G., Stasinska, G., Fullerton, A. W., Rauch, T. 2004, ApJ, 616, 485

\bibitem[1996]{TK1996}  Tweedy, R. W. \& Kwitter, K. B.  1996, ApJS, 107, 255

\bibitem[1993]{ty93} Tylenda, R., Acker, A., Stenholm, B. 1993, A\&AS, 102, 595

\bibitem[1993]{tyl}  Tylenda, R., \& G\'orny, S.K. 1993, AcA, 43, 389

\bibitem[1996]{WW1996} Walsh, J. R. \& Walton, N. A. 1996, A\&A, 315, 253

\bibitem[2008]{ww} Weidmann, W. A., Gamen, R., D\'iaz, R. J., Niemela, V. S. 2008, A\&A, 488, 245

\bibitem[1997]{WK1997}  Weinberger, R., Kerber, F., Groebner, H.  1997, A\&A, 323, 963

\bibitem[2006]{wehe06} Werner, K. \& Herwig, F., 2006, PASP, 118, 183

\bibitem[1994]{WO1994} Wlodarczyk, K. \& Olszewski, P. 1994, AcA, 44, 407

\bibitem[1991]{zhkw} Zhang, C. Y. \& Kwok, S. 1991, A\&A, 250, 179

\bibitem[1990]{ZP1990}  Zijlstra, A., Pottasch, S., Bignell, C.  1990, A\&AS, 82, 273




%
\addtocounter{table}{1}
%
\end{thebibliography}
\end{document}


\addtolength{\textwidth}{0cm}
\addtolength{\marginparwidth}{-4cm}
\addtolength{\voffset}{3cm}

{\small

\pagestyle{empty}

\begin{landscape}
\begin{longtable}{p{2cm}p{3.1cm}p{2.5cm}p{2.5cm}p{3cm}p{2cm}p{2cm}}
\caption{\label{ample2} 
Catalogue of CSPN (true PN)}\\
\hline\hline
  PN G  &  name  &  RA  &  DEC  &  classif.  &  ref. &  binary ref.   \\
\hline
\endfirsthead
\hline\hline
  PN G  &  name  &  RA  &  DEC  &  classif.  &  ref.  &  binary ref.   \\
\hline
\endhead
\hline
\endfoot
 000.0-06.8 & H 1-62 & 18 13 17.9 & -32 19 43.0 & emission-line & t.w. &  \\
 000.1+17.2 & PC 12 & 16 43 49.3 & -18 56 33.0 &  OB & t.w. &  \\
 000.3+12.2 & IC 4634 & 17 01 33.6 & -21 49 32.8 & emission-line & t.w. &  \\
 000.4+04.4  & K 5-1 & 17 29 52.4 & -26 11 14.0 & wels & GS2004 &  \\
 000.4-01.9 & M 2-20 & 17 54 25.4 & -29 36  08.2 & [WC5-6] & AN2003 &  \\
 000.7+08.0 & MPA 1717-2356 & 17 17 09.0 & -23 56 29.0 & Blue & MASH-II & \\
 000.7+04.7 & H 2-11 & 17 29 25.9 & -25 49  06.6 & wels & GS2004 &  \\
 000.7-02.7 & M 2-21 & 17 58 09.6 & -29 44 20.1 & wels & GC2009 &  \\
 000.9-02.0 & Bl 2-13 & 17 56 02.8 & -29 11 16.2 & wels & GS2004 &  \\
 001.2-03.0 & H 1-47 & 18 00 37.6 & -29 21 50.5 & [WC11]? & GS2004 &  \\
 001.5-06.7 & SwSt 1 & 18 16 12.3 & -30 52  08.1 & [WC9]pec & AN2003  &  \\
 001.7-04.4 & H 1-55 & 18 07 14.6 & -29 41 24.5 & [WC11] & GC2009 &  \\
 001.7-04.6 & H 1-56 & 18 07 53.9 & -29 44 34.3 & wels & GC2009 &  \\
 002.0-06.2 & M 2-33 & 18 15 06.6 & -30 15 33.3 & O5f(H); \newline  wels & HP2007; \newline  AN2003 &  \\
 002.0-13.4 & IC 4776 & 18 45 50.6 & -33 20 32.0 & wels  & TA1993 &  \\
 002.1-02.2 & M 3-20 & 17 59 19.4 & -28 13 48.2 & Of; \newline  wels? & AK1987; \newline  GC2009 &  \\
 002.2-02.7 & M 2-23 & 18 01 42.6 & -28 25 44.2 & Of & AK1987 &  \\
 002.2-06.3 & H 1-63 & 18 16 18.5 & -30 07 35.8 & OB? & t.w. &  \\
 002.2-09.4 & Cn 1-5 & 18 29 11.7 & -31 29 59.2 & [WO4]pec & AN2003 &  \\
 002.4+05.8 & NGC 6369 & 17 29 20.5 & -23 45 34.8 & [WO3] & AN2003  &  \\
 002.4-03.7 & M 1-38 & 18 06 05.8 & -28 40 29.3 &  O; \newline [WC11]  &  HP2007 \newline GC2009 &  \\
 002.6+05.5 & K 5-3 & 17 30 41.2 & -23 45  00.4 & [WC4]  & GS2004 &  \\
 002.6+08.1 & H 1-11 & 17 21 17.7 & -22 18 35.1 & wels & AN2003 &  \\
 002.6-03.4  & M 1-37 & 18 05 25.8 & -28 22  04.3 & [WC11]?; \newline  peculiar & GZ2006; \newline  HP2007 &  \\
 002.7-52.4 & IC 5148 & 21 59 35.2 & -39 23  08.0 &  hgO(H) &  SECGPN &  \\
 003.1+02.9 & Hb 4 & 17 41 52.8 & -24 42  08.1 & [WO3] & AN2003 &  \\
 003.2-04.4 & KFL 12 & 18 10 30.8 & -28 19 22.9 & wels & GC2009 &  \\
 003.3-04.6 & Ap 1-12 & 18 11 35.1 & -28 22 36.6 & [WC11]? & GC2009 &  \\
 003.4-04.8 & H 2-43 & 18 12 48.0 & -28 19 59.7 & symbiotic star?; \newline  wels? & BM2000; \newline  GC2009 &  \\
 003.6+03.1 & M 2-14 & 17 41 57.3 & -24 11 16.1 & wels & GS2004 &  \\
 003.7-04.6 & M 2-30 & 18 12 34.4 & -27 58 11.6 & wels & GC2009 &  \\
 003.9-14.9 & Hb 7 & 18 55 38.0 & -32 15 47.1 & O3; \newline  wels & GP2001; \newline  TA1993 &  \\
 004.0-03.0  & M 2-29 & 18 06 40.9 & -26 54 56.0 & H-poor & PT1992 &  \\
 004.2-04.3 & H 1-60 & 18 12 25.2 & -27 29 12.8 & wels & TA1993    &  \\
 004.6+06.0 & He 2-244 & 17 33 37.6 & -21 46 25.0 & wels & AN2003  &  \\
 004.8-22.7 & He 2-436 & 19 32 06.7 & -34 12 57.5 & [WC4] & AN2003 &  \\
 004.9+04.9 & M 1-25 & 17 38 30.3 & -22 08 38.8 & [WC5-6]  & AN2003 &  \\
 004.9-04.9 & M 1-44 & 18 16 17.4 & -27 04 32.5 & symbiotic star?; \newline K2III & 
BM2000; \newline SECGPN &  bc-CSPN \\
 005.9-02.6 & MaC 1-10 & 18 09 12.9 & -25 04 33.3 & [WC8] & GS2003 &  \\
 006.0-03.6 & M 2-31 & 18 13 16.1 & -25 30  05.3 & [WC4] & AN2003  &  \\
 006.0-41.9 & PaRu 1-1       & 21 05 53.6 & -37 08 40.3 & O(H) & PT1992 &  \\
 006.3+04.4 & H 2-18 & 17 43 28.8 & -21 09 51.3 & Of? & AK1987 &  \\
 006.4+02.0 & M 1-31 & 17 52 41.4 & -22 21 57.0 & wels & TA1993  &  \\
 006.5-03.1 & H 1-61 & 18 12 34.0 & -24 50  00.5 & wels & GS2004 &  \\
 006.7-02.2 & M 1-41 & 18 09 29.9 & -24 12 23.5 & (WN) ? & SECGPN &  \\
 006.8+04.1 & M 3-15 & 17 45 31.7 & -20 58  01.8 & [WC4] & AN2003 &  \\
 006.8-19.8 & WRAY 16-423 & 19 22 10.6 & -31 30 38.7 & [WC4-6]/wels & GZ2003 &  \\
 006.9-05.1 & MPA 1820-2524 & 18 20 57.7 & -25 24 22 & Blue & MASH-II & \\
 007.0-06.8 & Vy 2-1 & 18 27 59.6 & -26 06 48.3 & wels & TA1993 &  \\
 007.5-05.0 & BMP 1822-2449 & 18 22 10.4 & -24 49 54.0 & Blue? & MASH-II & \\
 007.8-03.7 & M 2-34 & 18 17 15.9 & -23 58 54.5 & [WC] & SECGPN &  \\
 007.8-04.4 & H 1-65 & 18 20 08.9 & -24 15  05.0 & [WC11]; \newline  wels & GC2009; \newline  AN2003 &  \\
 008.0+03.9 & NGC 6445 & 17 49 15.2 & -20 00 34.5 & cont. & SECGPN &  \\
 008.1-04.7 & M 2-39 & 18 22 01.1 & -24 10 40.2 & wels & GS2004 &  \\
 008.2+06.8 & He 2-260 & 17 38 57.1 & -18 17 35.0 & O & HP2007 &  \\
 008.3-01.1 & M 1-40 & 18 08 26.0 & -22 16 53.3 & wels & GS2004 &  \\
 008.3-07.3 & NGC 6644 & 18 32 34.7 & -25 07 44.2 & wels & GS2004 &  \\
 009.4-05.0 & NGC 6629 & 18 25 42.4 & -23 12 10.6 & [WC4]?; \newline  wels & AN2003; \newline  TA1993 &  \\
 009.6+10.5 & A 41 & 17 29 02.0 & -15 13  04.4 & sdB & WO1994 & GB1983 \\
 009.6+14.8 & NGC 6309 & 17 14 04.3 & -12 54 37.7 & wels & GS2004 &  \\
 009.6-10.6 & M 3-33 & 18 48 12.1 & -25 28 52.4 & wels & GS2004 &  \\
 009.8-04.6 & H 1-67 & 18 25 05.0 & -22 34 52.6 & [WO2]  & GS2004 &  \\
 009.8-07.5 & IRAS 18333-2357 & 18 36 22.8 & -23 55 18.3 & sdO?; \newline  O(C) & HP1993; \newline  PT1992 &  \\
 010.4+04.4  & SAKURAI´S & 17 52 32.7 & -17 41  08.0 & born-again & DB1996 &  \\
 010.7-06.4 & IC 4732 & 18 33 54.7 & -22 38 41.0 & Of & AK1987 &  \\
 010.8+18.0  & M 2-9 & 17 05 38.0 & -10 08 34.6 & B[e] & LZ1998  &  \\
 010.8-01.8 & NGC 6578 & 18 16 16.5 & -20 27  02.7 & wels & TA1993 &  \\
 011.4+17.9 & DHW 1-2 & 17 06 55.0 & -09 46 59.0 & DAO & SW1997 &  \\
 011.7-00.6 & NGC 6567 & 18 13 45.2 & -19 04 34.2 & wels & TA1993 &  \\
 011.7-06.6 & M 1-55 & 18 36 42.5 & -21 48 59.4 & O? & SECGPN & \\
 011.9+04.2 & M 1-32 & 17 56 20.0 & -16 29  04.0 & [WO4]pec & AN2003 &  \\
 012.2+04.9 & PM 1-188 & 17 54 21.1 & -15 55 52.0 & [WC10] & AN2003 &  \\
 012.5-09.8 & M 1-62 & 18 50 26.1 & -22 34 22.8 & wels & TA1993 &  \\
 012.9+06.6 & BMP 1749-1429 & 17 49 39.7 & -14 29 18.0 & Blue? & MASH-II & \\
 013.7-10.6 & Y-C 2-32 & 18 55 30.6 & -21 49 39.0 & wels & GS2004 &  \\
 014.0-05.5 & V-V 3-5 & 18 36 32.3 & -19 19 28.0 & A & SECGPN & bc-CSPN \\
 014.2+03.8 & PMR 4 & 18 02 38.3 & -14 42  02.8 & wels & AN2003 &  \\
 014.3-05.5 & Sa 2-352 & 18 37 11.1 & -19 02 21.9 & wels & AN2003 &  \\
 014.4-06.1 & SB 19 & 18 39 40.1 & -19 14 12.0 & wels & GS2004 &  \\
 014.6-04.3 & M 1-50         & 18 33 20.9 & -18 16 37.1 & cont.? & SECGPN &  \\
 015.4-04.5 & M 1-53 & 18 35 48.2 & -17 36  08.4 & emission-line? & t.w. &  \\
 015.5+02.8 & BMP 1808-1406 & 18 08 35.1 & -14 06 43.0 & Blue & MASH-II & \\
 015.9+03.3  & M 1-39 & 18 07 30.7 & -13 28 47.6 & cont. & K1994 &  \\
 016.4-01.9 & M 1-46 & 18 27 56.3 & -15 32 54.6 & wels; \newline  Of(H) & TA1993; \newline  H2003 &  \\
 016.8-01.7 & BMP 1827-1504 & 18 27 50.8 & -15 04 24.0 & [WC4] & MASH-II &  \\
 017.3-21.9  & A 65 & 19 46 34.2 & -23 08 12.9 & Op k & SECGPN & WW1996 \\
 017.6-10.2 & A 51 & 19 01 01.4 & -18 12 15.3 & O(H) &  SECGPN &  \\
 017.9-04.8 & M 3-30 & 18 41 14.9 & -15 33 43.6 & [WO2] & AN2003 &  \\
 019.4-05.3 & M 1-61 & 18 45 55.1 & -14 27 37.9 & wels & AN2003 &  \\
 019.7-04.5 & M 1-60 & 18 43 38.1 & -13 44 48.9 & [WC4] & AN2003 &  \\
 019.7-10.7 & MPA 1906-1634 & 19 06 32.8 & -16 34 00.0 & Blue & MASH-II & \\
 020.4-07.0 & MPA 1854-1420 & 18 54 14.7 & -14 20 19.0 & Blue? & MASH-II & \\
 020.7-05.9 & Sa 1-8 & 18 50 44.2 & -13 31  02.4 & OB & t.w. &  \\
 020.7-08.0 & MPA 1858-1430 & 18 58 19.3 & -14 30 26.0 & Blue & MASH-II & \\ 
 020.9-01.1 & M 1-51 & 18 33 28.9 & -11 07 26.4 & [WO4]pec & AN2003 &  \\
 021.0-04.1 & PMR 7 & 18 44 07.7 & -12 26 51.0 & [WC4] & PM2003 &  \\
 023.8-06.2 & BMP 1857-1054 & 18 57 09.8 & -10 54 51.0 & Blue & MASH-II & \\ 
 025.3+40.8 & IC 4593 & 16 11 44.5 & +12 04 17.1 & O7; \newline  O5f(H) & 
BD1993; \newline  SECGPN &  \\
 025.4-04.7 & IC 1295 & 18 54 37.2 & -08 49 39.1 & hgO(H) & NS1995 &  \\
 025.8-17.9 & NGC 6818 & 19 43 57.8 & -14 09 11.9 & wels & TA1993 & BC2003 \\
 025.9-10.9 & NA 2 & 19 18 19.5 & -11 06 15.4 & wels & MA2003 &  \\
 027.6+04.2 & M 2-43 & 18 26 40.1 & -02 42 57.3 & [WC7-8] & AN2003 &  \\
 027.6+16.9 & DeHt 2 & 17 41 40.9 & +03 06 57.3 & O & SECGPN &  \\
 027.6-09.6 & IC 4846 & 19 16 28.2 & -09 02 36.5 & Of & SECGPN &  \\
 028.0+10.2 & WeSb 3 & 18 06 00.8 & +00 22 38.6 & PG 1159? & NS1995 &  \\
 029.2-05.9 & NGC 6751 & 19 05 55.6 & -05 59 32.9 & [WO4] & AN2003 &  \\
 030.6+06.2  & Sh 2-68 & 18 24 58.4 & +00 51 35.9 & hybrid; \newline  DAO & WH2006; \newline  P1996 &  \\
 033.2-01.9 & Sa 3-151 & 18 58 51.7 & -00 32 54.3 & A & SECGPN & bc-CSPN \\
 034.1-10.5 & HDW 11 & 19 31 07.2 & -03 42 31.5 & hgO(H) & NS1995 &  \\
 034.5-06.7 & NGC 6778 & 19 18 24.9 & -01 35 47.4 & cont. & F1994 & MC2010 \\
 034.6+11.8 & NGC 6572 & 18 12 06.4 & +06 51 13.0 & wels & TA1993 &  \\
 035.9-01.1  & Sh 2-71 & 19 01 59.3 & +02 09 18.0 & A7 V - F0 V & F1999 & F1999 \\
 036.0+17.6  & A 43 & 17 53 32.3 & +10 37 24.2 & hybrid & WH2006 &   \\
 036.1-57.1 & NGC 7293 & 22 29 38.6 & -20 50 13.6 & DAO & NS1995 &  \\
 036.4-01.9 & IRAS 19021+0209 & 19 04 38.5 & +02 14 23.0 & cont. & t.w. &  \\
 037.5-05.1 & A 58 & 19 18 20.5 & +01 46 59.6 & [WCE]  & SECGPN &  \\
 037.7-06.0 & MPA 1921+0132 & 19 21 44.5 & +01 32 40.0 & [WO3-4] & MASH-II &  \\
 037.7-34.5 & NGC 7009 & 21 04 10.9 & -11 21 48.3 & O(H) & SECGPN &  \\
 037.8-06.3 & NGC 6790 & 19 22 57.0 & +01 30 46.6 & WN? & SECGPN &  \\
 038.2+12.0 & Cn 3-1 & 18 17 34.1 & +10 09 03.3 & wels & TA1993 &  \\
 042.5-14.5 & NGC 6852 & 20 00 39.2 & +01 43 40.1 & PG 1159  & WH2006 &  \\
 042.9-06.9 & NGC 6807 & 19 34 33.5 & +05 41  02.5 & Of & AK1987 &  \\
 043.1+03.8 & M 1-65 & 18 56 33.6 & +10 52  09.7 & O6; \newline  CIII and HeII emission & SECGPN \newline K1994  &  \\
 043.1+37.7 & NGC 6210 & 16 44 29.5 & +23 47 59.7 & O6 & SECGPN &  \\
 043.3+11.6 & M 3-27 & 18 27 48.3 & +14 29  06.1 & H$\alpha$  emission & MV1997 &  \\
 045.4-02.7  & Vy 2-2 & 19 24 22.2 & +09 53 56.7 & B[e] & LZ1998 &  \\
 045.7-04.5 & NGC 6804 & 19 31 35.1 &  +09 13 31.4 & O9 & SECGPN &  \\
 046.4-04.1 & NGC 6803 & 19 31 16.5 & +10 03 21.7 & wels & TA1993 &  \\
 046.8+03.8  & Sh 2-78 & 19 03 10.1 & +14 06 58.9 & PG 1159 & WH2006 &  \\
 047.0+42.4 & A 39 & 16 27 33.7 & +27 54 33.5 & hgO(H) & NS1995 &  \\
 048.7+01.9 & He 2-429 & 19 13 38.4 &  +14 59 19.1 & [WC4] & AN2003 &  \\
 049.4+02.4  & He 2-428 & 19 13 05.2 & +15 46 39.8 & O5 & RC2001 & SG2010 \\
 051.0+02.8 & IRAS 19127+1717 & 19 14 59.7 & +17 22 46.0 & ? + B9 V & SECGPN & bc-CSPN \\ 
 051.9-03.8 & M 1-73 & 19 41 09.3 &  +14 56 58.8 & wels & TA1993 &  \\
 052.2-04.0 & M 1-74 & 19 42 18.9 & +15 09  08.2 & WN b? & SECGPN &  \\
 052.5-02.9  & Me 1-1 & 19 39 09.8 &  +15 56 48.2 & K(1-2) II & PM2008 & bc-CSPN \\
 053.8-03.0 & A 63 & 19 42 10.4 & +17 05 14.5 & M4 V & SECGPN & A1966 \\
 054.1-12.1 & NGC 6891 & 20 15 08.8 & +12 42 15.6 & wels & TA1993 &  \\
 055.1-01.8 & K 3-43 & 19 40 25.9 & +18 49 14.2 & M & SF1987 & bc-CSPN \\
 055.4+16.0 & A 46 & 18 31 18.3 & +26 56 12.9 & M6 V; \newline  O9k & MO2007; \newline  SECGPN & MO2007 \\
 055.5-00.5 & M 1-71 & 19 36 26.9 & +19 42 24.1 & wels  & TA1993 &  \\
 057.2-08.9 & NGC 6879 & 20 10 26.7 & +16 55 21.4 & wels & AN2003 &  \\
 058.3-10.9 & IC 4997 & 20 20 08.8 & +16 43 53.6 & wels & TA1993 &  \\
 060.1-07.7  & NGC 6886 & 20 12 42.8 & +19 59 22.6 & cont. & CP1985 &  \\
 060.4+01.5 & PM 1-310 & 19 38 52.1 & +25 05 32.6 & [WC11] & SECGPN &  \\
 060.8-03.6 & NGC 6853 & 19 59 36.3 & +22 43 16.1 & DAO & NS1995 &  \\
 061.0+08.0 & K 3-27 & 19 14 30.0 & +28 40 45.5 & G0: & L1977 & bc-CSPN \\
 061.4-09.5 & NGC 6905 & 20 22 22.9 & +20 06 16.8 & [WO2] & AN2003 &  \\
 061.8+02.1 & He 2-442 & 19 39 43.4 & +26 29 33.1 & symbiotic star? & BM2000 &  \\
 061.9+41.3  & DDDM 1         & 16 40 18.2 & +38 42 20.0 & O(H) & PT1992 &  \\
 062.4+09.5 & NGC 6765 & 19 11 06.5 & +30 32 42.5 & PG 1159 & WH2006 &  \\
 063.1+13.9 & NGC 6720 & 18 53 35.1 & +33 01 45.0 & DA(O?) & NS1995  &  \\
 064.6+48.2 & NGC 6058 & 16 04 26.6 & +40 40 56.1 & O9 & P1983  &  \\
 064.7+05.0 & BD+30 3639 & 19 34 45.2 & +30 30 58.9 & [WC9] & AN2003 &  \\
 065.0-27.3  & Ps 1           & 21 29 59.4 & +12 10 27.5 & sdO; \newline  O(H)-C & RH2002; \newline  PT1992 &  \\
 066.7-28.2  & NGC 7094 & 21 36 53.0 &  +12 47 19.0 & hybrid  & WH2006 &  \\
 068.3-02.7 & He 2-459 & 20 13 57.9 & +29 33 55.9 & [WC9] & AN2003 &  \\
 068.7+14.8 & SP 4-1 & 19 00 26.6 & +38 21  07.3 & wels & TA1993 &  \\
 069.4-02.6 & NGC 6894 & 20 16 24.0 & +30 33 53.2 & WD? & SZ1997 &  \\
 072.7-17.1 & A 74 & 21 16 52.3 & +24 08 51.8 & DAO & NS1995 &  \\
 075.7+35.8 & Sa 4- 1 & 17 13 50.4 & +49 16 11.0 & O(H) & SECGPN &  \\
 077.6+14.7 & A 61 & 19 19 10.2 & +46 14 52.0 & hgO(H) & NS1995 &  \\
 080.3-10.4 & RX J2117.1+3412 & 21 17 08.3 & +34 12 27.5 & PG 1159 & WH2006 &  \\
 081.2-14.9 & A 78 & 21 35 29.4 & +31 41 45.3 & [WC]-PG1159 & GT2000 &  \\
 082.1+07.0 & NGC 6884 & 20 10 23.7 & +46 27 39.8 & WN b? & SECGPN &  \\
 082.5+11.3 & NGC 6833 & 19 49 46.6 & +48 57 40.2 & Of & AK1987 &  \\
 083.5+12.7 & NGC 6826 & 19 44 48.2 & +50 31 30.3 & O3f(H) & SECGPN & LR1983 \\
 089.0+00.3 & NGC 7026 & 21 06 18.2 & +47 51  05.4 & [WO3] & AN2003 &  \\
 089.3-02.2 & M 1-77 & 21 19 07.4 & +46 18 47.2 & OB ? & SECGPN &  \\
 089.8-05.1 & IC 5117 & 21 32 31.0 &  +44 35 48.5 & [WR] & SECGPN &  \\
 093.4+05.4 & NGC 7008 & 21 00 32.5 & +54 32 36.2 & O7 & SECGPN & CB1999 \\
 093.9-00.1 & IRAS 21282+5050 & 21 29 58.4 & +51 03 59.8 & O7(f)-[WC11] & CJ1987 &  \\
 094.0+27.4 & K 1-16 & 18 21 52.2 & +64 21 54.3 & PG 1159 (lg E); \newline  [WO] & WH2006; \newline  SK1994 & FS2010 \\
 095.2+00.7 & K 3-62         & 21 31 50.2 & +52 33 51.6 & cont. & SECGPN &  \\
 096.3+02.3 & K 3-61 & 21 30 00.7 & +54 27 27.5 & [WC4-6] & TA1993 &  \\
 096.4+29.9 & NGC 6543 & 17 58 33.4 & +66 37 59.5 & wels & TA1993 &  \\
 096.8+31.9 & RE 1738+665 & 17 38 02.6 & +66 53 47.9 & DA & N1999 &  \\
 096.9+32.0 & RE 1738+665 & 17 37 59.1 & +66 53 46.3 & DA & TK1996 &  \\
 100.0-08.7 & Me 2-2 & 22 31 43.7 & +47 48  03.9 & Of & AK1987 &  \\
 100.6-05.4 & IC 5217 & 22 23 55.7 &  +50 58 00.5 & [WC8-9]?; \newline  wels & AN2003; \newline  TA1993 &  \\
 102.9-02.3  & A 79 & 22 26 17.3 & +54 49 38.2 & F0 V & B2008 & bc-CSPN \\
 103.7+00.4 & M 2-52 & 22 20 30.7 & +57 36 21.6 & Non-[WC] & PM2002 &  \\
 104.2-29.6 & Jn 1 & 23 35 53.3 & +30 28  06.4 & PG 1159 (E) & WH2006 &  \\
 104.4-01.6 & M 2-53 & 22 32 17.7 & +56 10 26.1 & WN a? & SECGPN &  \\
 104.8-06.7 & M 2-54 & 22 51 38.9 & +51 50 42.4 & B & SECGPN & bc-CSPN \\
 106.5-17.6 & NGC 7662 & 23 25 53.6 & +42 32  06.0 & UV emission lines & F1994 &  \\
 107.7+07.8 & IsWe 2 & 22 13 22.5 & +65 53 55.5 & DA & NS1995 &  \\
 107.7-02.2 & M 1-80         & 22 56 19.8 & +57 09 20.7 & cont. & SECGPN &  \\
 107.8+02.3 & NGC 7354 & 22 40 19.8 & +61 17  08.7 & cont. & F1994 &  \\
 111.0+11.6 & DeHt 5 & 22 19 33.7 & +70 56  03.1 & DA & NS1995 &  \\
 111.8-02.8  & Hb 12 & 23 26 14.8 & +58 10 54.7 & B[e]?; \newline WN7? & 
LZ1998; \newline SECGPN & HI2006 \\
 114.0-04.6  & A 82 & 23 45 47.8 & +57 03 58.5 & K0 IV & CB1999 & bc-CSPN \\
 118.0-08.6 & Vy 1-1 & 00 18 42.2 & +53 52 20.0 & O(H); \newline  [WC] & N1999; \newline  AK1985 &  \\
 118.8-74.7 & NGC 246 & 00 47 03.3 & -11 52 18.9 & PG 1159 (lg E) & WH2006 & BC1999 \\
 119.6-06.7 & Hu 1-1 & 00 28 15.6 & +55 57 54.7 & A? & SECGPN & bc-CSPN \\
 120.0+09.8 & NGC 40 & 00 13 01.0 & +72 31 19.1 & [WC8] & AN2003 &  \\
 120.2-05.3 & Sh 2-176 & 00 31 53.3 & +57 22 49.0   & DA & NS1995 &  \\
 120.3+18.3 & Sh 2-174 & 23 45 02.3 & +80 56 59.6 & DAO & NS1995 &  \\
 123.6+34.5 & IC 3568 & 12 33 06.9 & +82 33 49.0 & O3(H); \newline  O5f & SECGPN; \newline  P1983 &  \\
 124.0+10.7 & EGB 1 & 01 07 07.6 & +73 33 23.1 & DA & NS1995 &  \\
 125.9-47.0 & PHL 932 & 00 59 56.7 & +15 44 13.8 & hgO(H) & SECGPN &  \\
 126.6+01.3 & PRINCIPES \newline DE ASTURIAS & 01 25 08.0 & +63 56 52.7 &  H$\alpha$ and CaII emission & MC2006 & MC2006 \\
 128.0-04.1 & Sh 2-188 & 01 30 33.1 & +58 24 50.7 & DAO & NS1995 &  \\
 130.2+01.3 & IC 1747 & 01 57 35.9 & +63 19 19.4 & [WO4] & AN2003  &  \\
 130.9-10.5 & NGC 650-1 & 01 42 19.9 & +51 34 31.2 & PG 1159 (E) & WH2006 &  \\
 135.6+01.0 & WeBo 1 & 02 40 14.4 & +61 09 16.8 & K0 III & SB2007 & BP2003 \\
 135.9+55.9 & TS 01 & 11 53 24.7 &  +59 39 56.9 & WD/NS & TN2004 & NT2005 \\
 136.3+05.5 & HFG 1 & 03 03 47.0 & +64 54 35.7 & F9 V & EP2005 & GB1987 \\
 138.1+04.1 & HDW 2 & 03 11 01.3 & +62 47 45.1 & A & SECGPN & bc-CSPN \\
 138.8+02.8 & IC 289 & 03 10 19.3 & +61 19  01.0 & O(H); \newline  WD? & N1999; \newline  CG2009 &  \\
 143.6+23.8 & EGB 4 & 06 29 34.0 & +71 04 36.3 & H-rich & NS1995 & NS1995 \\
 144.5+06.5 & NGC 1501 & 04 06 59.2 & +60 55 14.3 & [WO4] & AN2003 &  \\
 144.8+65.8  & BE Uma & 11 57 44.8 & +48 56 18.7 & M3 V & MD1981 & FM1981 \\
 146.7+07.6 & M 4-18 & 04 25 50.9 & +60 07 12.8 & [WC11] & AN2003 &  \\
 147.4-02.3 & M 1-4 & 03 41 43.4 & +52 17  00.3 & Of & AK1987 &  \\
 148.4+57.0 & NGC 3587 & 11 14 47.7 & +55 01  08.5 & hgO(H) & NS1995 &  \\
 149.4-09.2 & HDW 3 & 03 27 15.4 & +45 24 20.5 & DAO & NS1995 &  \\
 149.7-03.3 & IsWe 1 & 03 49 05.9 & +50 00 14.8 & PG 1159 (A) & GT2000 &  \\
 156.3+12.5 & HDW 4 & 05 37 56.2 & +55 32 16.0 & DA & NS1995 &  \\
 156.9-13.3 & HaWe 5 & 03 45 26.6 & +37 48 51.8 & DA & NS1995 &  \\
 158.6+00.7 & Sh 2-216 & 04 43 21.3 & +46 42  05.8 & DAO & NS1995 &  \\
 158.9+17.8 & PuWe 1 & 06 19 34.3 & +55 36 42.3 & DAO & NS1995 &  \\
 159.0-15.1 & IC 351 & 03 47 33.1 & +35 02 48.5 & wels &  dS2003 &  \\
 161.2-14.8 & IC 2003 & 03 56 22.0 & +33 52 30.6 & [WC3]?; \newline [WC7-8] & TA1993; \newline SECGPN &  \\
 164.8+31.1 & VV 47 & 07 57 51.6 & +53 25 17.0 & PG 1159 (E:) & WH2006 &  \\
 165.5-06.5  & K 3-67 & 04 39 47.9 & +36 45 42.6 & O(C); \newline CIV+HeII emission & SECGPN \newline TS1987 &  \\
 165.5-15.2 & NGC 1514 & 04 09 17.0 & +30 46 33.5 & sdO + A0 III & SECGPN & bc-CSPN \\
 166.1+10.4 & IC 2149 & 05 56 23.9 & +46 06 17.3 & O4f & SECGPN &  \\
 167.4-09.1 & K 3-66         & 04 36 37.2 & +33 39 30.0 & cont. & SECGPN &  \\
 169.6+00.0 & IC 2120 & 05 18 10.3 & +37 33 27.4 & G & ZP1990 & bc-CSPN \\
 170.3+15.8 & NGC 2242 & 06 34 07.4 & +44 46 38.1 & O(H) & PT1992 &  \\
 189.1+19.8 & NGC 2371-72 & 07 25 34.7 &  +29 29 26.4 & [WO1] & AN2003 &  \\
 189.8+07.7 & M 1-7 & 06 37 21.0 & +24 00 35.4 & [WO]-[WC8] & SECGPN &  \\
 190.3-17.7 & J 320 & 05 05 34.3 & +10 42 22.7 & wels & TA1993 &  \\
 191.4+33.1 & Ton 320 & 08 27 05.5 & +31 30  08.6 & DAO & N1999 &  \\
 193.6-09.5 & H 3-75 & 05 40 45.0 & +12 21 23.3 & G-K & BP2002 & bc-CSPN \\
 194.2+02.5 & J 900 & 06 25 57.3 & +17 47 27.2 & wels & TA1993 &  \\
 196.6-10.9 & NGC 2022 & 05 42 06.2 & +09 05 10.3 & O(H) & N1999 & \\
 197.4-06.4 & WeDe 1 & 05 59 24.9 & +10 41 40.4 & DA & NS1995 &  \\
 197.8+17.3 & NGC 2392 & 07 29 10.8 & +20 54 42.5 & O6f & SECGPN & CB1999 \\
 197.8-03.3 & A 14 & 06 11 08.7 & +11 46 43.8 & B5 III-V & LK1987 & bc-CSPN \\
 204.1+04.7 & K 2-2 & 06 52 23.2 & +09 57 55.7 & hgO(H) & NS1995 &  \\
 205.1+14.2 & A 21 & 07 29 02.7 & +13 14 48.4 & PG 1159 (E) & WH2006 &  \\
 206.4-40.5 & NGC 1535 & 04 14 15.8 & -12 44 22.0 & O5 & SECGPN & CB1999 \\
 208.5+33.2 & A 30 & 08 46 53.5 & +17 52 45.5 & [WC]-PG1159 & GT2000 & CB1999 \\
 211.2-03.5 & M 1-6 & 06 35 45.1 & -00 05 37.4 & emission-line & t.w. &  \\
 211.4+18.4 & HDW 7 & 07 55 11.3 & +09 33  09.3 & hgO(H) & SECGPN &  \\
 211.9+22.6 & EGB 5 & 08 11 12.8 & +10 57 17.1 & hgO(H) & SECGPN &  \\
 214.9+07.8 & A 20 & 07 22 57.7 & +01 45 32.8 & O(H) & SECGPN &  \\
 215.2-24.2 & IC 418 & 05 27 28.2 & -12 41 50.3 & Of(H) & SECGPN &  \\
 215.5-30.8 & A 7 & 05 03 07.5 & -15 36 22.7 & DAO &  NS1995 & CB1999 \\
 215.6+03.6 & NGC 2346 & 07 09 22.5 & -00 48 23.6 & A5 V & SECGPN & MN1981 \\
 215.7-03.9 & BMP 0642-0417 & 06 42 18.4 & -04 17 49.0 & Blue & MASH-II & \\
 216.0+07.4 &PHR 0723+0036 & 07 23 48.1 & +00 36 48.0 & [WR] & MASH-I   &      \\
 218.9-10.7 & HDW 5 & 06 23 37.2 & -10 13 23.7 & hgO(H); \newline unknown absorption at 5758\AA. & SECGPN \newline LS2007 &  \\
 219.1+31.2 & A 31 & 08 54 13.2 & +08 53 53.1 & hgO(H) & NS1995 & CB1999 \\
 219.5+02.8 & BMP 0713-0432 & 07 13 51.0 & -04 32 51.0 & Blue? & MASH-II & \\
 220.3-53.9  & NGC 1360 & 03 33 14.6 & -25 52 18.0 & O(H); \newline  WD/MS & SECGPN; \newline  d2006 & MN1977 \\
 221.0-01.4 & PHR 0701-0749 & 07 01 09.3 & -07 49 21.0 & wels  & MASH-I  &   \\
 221.3-12.3 & IC 2165 & 06 21 42.8 & -12 59 14.0 & wels & AN2003 &  \\
 221.5+46.3 & EGB 6 & 09 52 59.0 & +13 44 34.9 & hgO(H) & SECGPN & BM1993 \\
 222.1+03.9 & PFP 1 & 07 22 17.7 & -06 21 46.0 & pre-WD? & PF2004 &  \\
 222.5+07.6 & BMP 0736-0500 & 07 36 23.1 & -05 00 20.0 & Blue & MASH-II & \\ 
 222.8-04.2 & PM 1-23 & 06 54 13.4 & -10 45 38.0 & [WC7] & SG2006 & HZ2010 \\
 224.9+01.0 & We 1-6 & 07 17 26.0 & -10 10 37.7 & hgO(H) & SECGPN &  \\
 225.5-02.5 & MPA 0705-1224 & 07 05 37.2 & -12 24 52.0 & Blue & MASH-II & \\
 228.0-00.4 & MPA 0717-1334 & 07 17 57.5 & -13 34 08.0 & Blue? & MASH-II & \\
 228.2-22.1 & DeHt 1 & 05 55 06.7 & -22 54  02.2 & K V & d2006 & bc-CSPN \\
 231.1+03.9 & BMP 0739-1418 & 07 39 50.6 & -14 18 26.0 & Blue & MASH-II & \\
 231.8+04.1  & NGC 2438 & 07 41 50.5 & -14 44  07.7 & M3 V; \newline  O(H) & d2009; \newline  N1999 & BC2008 \\
 232.0+05.7 & SaSt 2-3 & 07 48 03.5 & -14 07 42.6 & OB & t.w. &  \\
 232.8-04.7 & M 1-11 & 07 11 16.7 & -19 51  02.9 & emission-line & t.w. &  \\
 233.5-16.3 & A 15 & 06 27 01.9 & -25 22 49.6 & O(H) & SECGPN &  \\
 234.3-07.2 & MPA 0704-2221 & 07 04 23.0 & -22 21 52.0 & Blue & MASH-II & \\
 234.9-01.4 & M 1-14 & 07 27 56.5 & -20 13 23.4 & OB & t.w. &  \\
 234.9-09.7 & MPA 0656-2356 & 06 56 00.0 & -23 56 49.0 & Blue & MASH-II & \\
 235.3-03.9 & M 1-12 & 07 19 21.5 & -21 43 55.4 & emission-line & t.w. &  \\
 237.3-08.4 & BMP 0705-2528 & 07 05 45.5 & -25 28 50.0 & Blue & MASH-II & \\
 237.4-09.6 & BMP 0700-2607 & 07 00 51.8 & -26 07 18.0 & Blue & MASH-II & \\ 
 238.0+34.8 & A 33 & 09 39 09.1 & -02 48 32.0 & O(H) & SECGPN & CB1999 \\
 239.6+13.9 & NGC 2610 & 08 33 23.3 & -16 08 57.7 & WD? & CG2009 & CB1999 \\
 240.3+07.0 & Y-C 2-5 & 08 10 41.7 & -20 31 32.9 & emission-line & t.w. &  \\
 240.8-19.6 & KLSS 1-9 & 06 24 36.4 & -33 04 49.0 & OB & t.w. &  \\
 241.0+02.3 & M 3-4 & 07 55 11.2 & -23 37 45.6 & cont. & t.w. &  \\
 242.6-11.6 & M 3-1 & 07 02 49.6 & -31 35 41.3 & cont. & t.w. &  \\
 243.3-01.0 & NGC 2452 & 07 47 26.3 & -27 20  06.6 & [WO1] & AN2003 &  \\
 243.8-37.1 & PRTM 1 & 05 03 01.7 & -39 45 44.5 & O(H) & SECGPN &  \\
 245.0+02.2 & BMP 0803-2706 & 08 03 54.2 & -27 06 02.0 & Blue & MASH-II & \\
 245.1-05.5 & BMP 0733-3108 & 07 33 24.1 & -31 08 05.0 & Blue & MASH-II & \\
 245.4+01.6 & M 3-5 & 08 02 28.9 & -27 41 55.4 & O7 & SECGPN &  \\
 248.7+29.5 & A 34 & 09 45 35.4 & -13 10 15.8 & hgO(H) & SECGPN &  \\
 248.8-08.5 & M 4-2 & 07 28 55.2 & -35 45 15.4 & emission-line & t.w. &  \\
 250.5+01.9 & BMP 0816-3150 & 08 16 20.8 & -31 51 00.0 & Blue & MASH-II & \\
 250.6+09.3 & BMP 0844-2737 & 08 44 37.9 & -27 37 15.0 & Blue & MASH-II & \\
 252.6+04.4 & K 1-1 & 08 31 52.6 & -32 06  08.7 & G-K & SECGPN & bc-CSPN \\
 253.5+10.7  & K 1-2 & 08 57 46.0 & -28 57 36.8 & K2 V \newline (earlier than) & d2009 & BG1987 \\
 253.9+05.7 & M 3-6 & 08 40 40.2 & -32 22 33.6 & wels & AN2003 &  \\
 254.6+00.2 & Ns 238 & 08 20 56.7 & -36 13 46.7 & OB & t.w. &  \\
 255.3-59.6 & Lo 1 & 02 56 58.4 & -44 10 17.8 & hgO(H) & SECGPN &  \\
 257.5+00.6  & VBRC 1 & 08 30 54.2 & -38 18  07.0 & F V: & RK1999 & bc-CSPN \\
 258.0-03.2 & BMP 0815-4053 & 08 15 56.9 & -40 53 08.0 & Blue & MASH-II & \\
 258.0-15.7 & wray 17-1 & 07 14 49.4 & -46 57 39.1 & PG 1159 & WH2006 &  \\
 258.1-00.3 & He 2-9 & 08 28 28.0 & -39 23 40.3 & wels & TA1993 &  \\
 258.5-01.3 & RCW 24 & 08 25 47.5 & -40 13 10.3 & absorption lines & FP2006 &  \\
 261.0+32.0 & NGC 3242 & 10 24 46.1 & -18 38 32.6 & O(H) & SECGPN &  \\
 261.9+08.5 & NGC 2818 & 09 16 01.7 & -36 37 38.8 & cont. & LS2007 &  \\
 261.9-05.3 & BMP 0818-4517 & 08 18 16.8 & -45 17 57.0 & Blue & MASH-II & \\
 263.0-05.5 & PB 2 & 08 20 39.8 & -46 20 13.2 & emission-line? & t.w. &  \\
 263.1+04.3 & FPM 0904-4023 & 09 04 02.3 & -40 22 20.0 & Blue & MASH-II & \\
 263.2+00.4 & K 2-15 & 08 48 47.7 & -42 54 24.0 & O(H) & SECGPN &  \\
 264.4-12.7 & He 2-5 & 07 47 20.0 & -51 15  03.4 & wels & AN2003 &  \\
 264.5+05.0 & FPM 0911-4051 & 09 11 45.6 & -40 51 59.0 & Blue & MASH-II & \\
 264.6+03.8 & BMP 0907-4146 & 09 07 24.3 & -41 46 14.0 & Blue & MASH-II & \\
 272.1+12.3 & NGC 3132 & 10 07 01.8 & -40 26 11.1 & A2 V & SECGPN & CB1999 \\
 272.8+01.0 & PMR 1 & 09 28 40.6 & -49 36 44.0 & [WC9-10] & AN2003 &  \\
 273.6+06.1 & HBDS 1 & 09 52 44.5 & -46 13 51.0 & O(H) & SECGPN  &  \\
 274.3+09.1 & Lo 4 & 10 05 45.8 & -44 21 33.3 & PG 1159 & WH2006  &  \\
 274.6+02.1 & He 2-35 & 09 41 37.5 & -49 57 58.6 & wels & TA1993 &  \\
 275.0-04.1 & PB 4 & 09 15 07.6 & -54 52 38.5 & emission-line? & t.w. &  \\
 275.2-03.7  & He 2-25 & 09 18 01.3 & -54 39 29.2 & symbiotic star?; \newline  cont. & C1995; \newline  LS2007 &  \\
 276.1-11.9 & MPA 0835-6039 & 08 35 07.3 & -60 39 43.0 & Blue & MASH-II & \\
 277.1-03.8  & NGC 2899 & 09 27 03.0 & -56 06 21.1 & F V: & RK1999 & bc-CSPN \\
 277.7-3.5 & VBRC 2 & 09 31 20.5 & -56 17 39.4 & DAO & PR1997 &  \\
 278.1-05.9 & NGC 2867 & 09 21 25.3 & -58 18 40.7 & [WO2] & AN2003 &  \\
 278.8+04.9 & PB 6 & 10 13 15.9 & -50 19 59.3 & [WO1] & AN2003 &  \\
 279.6-03.1 & He 2-36 & 09 43 25.6 & -57 16 55.6 & A2 III; \newline sdO + A2 III & 
LS2007; \newline SECGPN & bc-CSPN \\
 280.1-05.1 & BMP 0936-5905 & 09 36 43.6 & -59 05 17.0 & Blue? & MASH-II & \\
 281.0-05.6 & IC 2501 & 09 38 47.5 & -60 05 27.9 & emission-line & t.w. &  \\
 283.6+25.3 & K 1-22 & 11 26 43.8 & -34 22 11.4 & F V & RK1999 & CB1999 \\
 283.9+09.7 & DS 1 & 10 54 40.6 & -48 47  03.0 & M5 V & SECGPN & D1985 \\
 284.2-05.3 & PM 12 & 10 01 18.9 & -61 52  01.0 & [WR] & MASH-I &  \\
 285.4+01.5 & Pe 1-1 & 10 38 27.6 & -56 47  06.5 & [WO4] & AN2003 &  \\
 285.4-05.3 & IC 2553 & 10 09 21.7 & -62 36 40.9 & emission-line & t.w. &  \\
 285.6-02.7 & He 2-47 & 10 23 09.0 & -60 32 34.3 & emission-line & t.w. &  \\
 285.7-14.9 & IC 2448 & 09 07 06.3 & -69 56 30.7 & O(H) & SECGPN &  \\
 286.3+02.8 & He 2-55 & 10 48 43.2 & -56 03 10.2 & [WO3] & AN2003 &  \\
 286.8-29.5  & K 1-27 & 05 57 02.2 & -75 40 22.7 & O(He) & RD1998 & CB1999 \\
 289.6-01.6  & He 2-57 & 10 56 03.0 & -61 28  07.4 & symbiotic star? & KB1994 &  \\
 289.8+07.7 & He 2-63 & 11 24 01.0 & -52 51 19.4 & wels & AN2003 &  \\
 291.3+08.4 & PMR 2 & 11 34 38.6 & -52 43 33.0 & [WO4]-[WC4] & AN2003 &  \\
 291.3-26.2 & Vo 1 & 06 59 26.4 & -79 38 47.0 & [WC10] & AN2003 &  \\
 291.4+19.2 & LoTr 4 & 11 52 29.2 & -42 17 38.7 & O(He) & GT2000 &  \\
 291.6-04.8 & IC 2621 & 11 00 19.5 & -65 14 54.2 & emission-line & t.w. &  \\
 291.7+03.7 & He 2-64 & 11 27 24.3 & -57 17 58.9 & O(He) & LS2007 &  \\
 292.4+04.1 & PB 8 & 11 33 17.7 & -57 06 14.0 & [WC5-6] & AN2003  &  \\
 293.2-09.5 & MPA 1054-7013 & 10 54 27.3 & -70 13 12.0 & Blue & MASH-II & \\
 293.6+10.9 & BlDz 1 & 11 53 06.6 & -50 50 59.2 & H-rich & RK1999 &  \\
 294.1+14.4 & Lo 6 & 12 00 43.5 & -47 33 12.0 & cont. & t.w. &  \\
 294.1+43.6 & NGC 4361 & 12 24 30.8 & -18 47  05.4 & O6 & P1983 &  \\
 296.0-06.2 & MPA 1137-6806 & 11 37 15.6 & -68 06 45.0 & Blue & MASH-II & \\
 297.0+06.5 & BMP 1209-5553 & 12 09 29.1 & -55 53 34.0 & [WO1-2] & MASH-II &  \\
 300.1+04.1 & BMP 1229-5839 & 12 29 57.6 & -58 39 06.0 & Blue & MASH-II & \\
 300.7-02.0 & He 2-86 & 12 30 30.5 & -64 52  05.7 & [WC4] & AN2003 &  \\
 301.9-02.1 & MPA 1242-6459 & 12 42 24.2 & -64 59 25.0 & wels? & MASH-II &  \\
 302.0-01.6 & MPA 1243-6428 & 12 43 19.4 & -64 28  01.0 & [WC9] & MASH-II &  \\
 303.6+40.0 & A 35 & 12 53 32.8 & -22 52 22.6 & ? + G8 IV & SECGPN & bc-CSPN \\
 305.1+01.4  & He 2-90 & 13 09 36.2 & -61 19 36.0 & B[e] & KB2005 &  \\
 306.4-00.6 & Th 2-A & 13 22 33.8 & -63 21  01.3 & [WO3]pec & WG2008 &  \\
 307.2-03.4 & NGC 5189 & 13 33 32.9 & -65 58 27.1 & [WO1] & AN2003 &  \\
 307.2-05.3 & MPA 1337-6751 & 13 37 22.4 & -67 51 07.0 & Blue & MASH-II & \\
 307.2-09.0 & He 2-97 & 13 45 24.0 & -71 28 48.8 & emission-line & t.w. &  \\
 307.5-04.9 & MyCn 18 & 13 39 35.1 & -67 22 51.9 & Of(C) & LS2007 &  \\
 308.2+07.7 & MeWe 1-3 & 13 28 04.9 & -54 41 58.4 & DAO & SW1997 &  \\
 308.4+00.4 & WeKg 2 & 13 38 41.7 & -61 55 51.0 & H-rich & WK1997 &  \\
 308.5+02.5 & PMR 6 & 13 36 23.0 & -59 53 31.0 & [WC4] & PM2003 &  \\
 308.6-12.2 & He 2-105 & 14 15 25.7 & -74 12 49.8 & OB & t.w. &  \\
 309.0-04.2 & He 2-99 & 13 52 30.7 & -66 23 26.8 & [WC9] & AN2003 &  \\
 309.1-04.3 & NGC 5315 & 13 53 57.0 & -66 30 50.7 & [WO4]  & AN2003 &  \\
 309.6-04.8 & MPA 1400-6647 & 14 00 37.1 & -66 47 58.0 & Blue & MASH-II & \\
 309.8-01.6 & MPA 1354-6337 & 13 54 22.3 & -63 37 18.0 & [WO4]? & MASH-II &  \\
 310.3+24.7 & Lo 8 & 13 25 37.5 & -37 36 14.8 & O(H) & SECGPN &  \\
 311.0+02.4  & SuWt 2 & 13 55 43.2 & -59 22 39.9 & B9 V & MO2007 & BO2002 \\
 312.3+10.5 & NGC 5307 & 13 51 03.3 & -51 12 15.9 & emission-line & t.w. &  \\
 312.6-01.8 & He 2-107 & 14 18 42.5 & -63 07 10.7 & emission-line & t.w. &  \\
 313.4+06.2 & MPA 1405-5507 & 14 05 32.2 & -55 07 44.0 & [WC4-6] & MASH-II &  \\
 313.8-05.7 & BMP 1442-6615 & 14 42 46.6 & -66 15 02.0 & Blue & MASH-II & \\
 313.9+02.8 & PM 9 & 14 16 37.6 & -58 09 30.0 & [WR] & MASH-II &  \\
 315.1-13.0 & He 2-131 & 15 37 11.2 & -71 54 52.9 & wels & TA1993 &  \\
 316.1+08.4 & He 2-108 & 14 18 08.9 & -52 10 39.5 & wels & TA1993 &  \\
 316.7-05.8 & MPA 1508-6455 & 15 08 06.4 & -64 55 49.0 & Blue & MASH-II & M2010 \\
 318.4+41.4 & A 36 & 13 40 41.3 & -19 52 55.3 & O(H) & SECGPN &  \\
 319.6+15.7 & IC 4406 & 14 22 26.3 & -44 09  04.4 & [WR] & SECGPN &  \\
 320.1-09.6 & He 2-138 & 15 56 01.7 & -66 09  09.2 & Of(H); \newline O(H) & 
MK1988; \newline SECGPN &  \\
 320.3-28.8 & He 2-434 &  19 33 50.7 & -74 32 58.7 &  OB & t.w. &  \\
 320.7+04.2 & BMP 1457-5413 & 14 57 02.3 & -54 13 58.0 & Blue? & MASH-II & \\
 321.0+03.9 & He 2-113 & 14 59 53.5 & -54 18  07.5 & [WC10] & AN2003 &  \\
 322.5-05.2 & NGC 5979 & 15 47 40.6 & -61 13  02.7 & emission-line & t.w. &  \\
 323.9+02.4 & He 2-123 & 15 22 19.4 & -54 08 12.8 & wels? & LS2007 &  \\
 324.0+03.5 & PM 1-89 & 15 19 08.8 & -53 09 49.8 & [WO4]pec & AN2003 &  \\
 324.1+09.0 & ESO 223-10 & 15 01 40.7 & -48 21  02.0 & O ? & SECGPN &  \\
 325.8+04.5 & He 2-128 & 15 25 07.9 & -51 19 40.9 & emission-line? & t.w. &  \\
 325.8-12.8 & He 2-182 & 16 54 35.1 & -64 14 28.5 & O(H) & SECGPN &  \\
 326.0-06.5 & He 2-151 & 16 15 42.3 & -59 54  01.0 & O(H) & SECGPN &  \\
 326.6+05.7 & BMP 1525-4957 & 15 25 14.1 & -49 57 41.0 & Blue & MASH-II & \\
 326.9+08.2 & MPA 1518-4738 & 15 18 18.2 & -47 38 28.0 & [WR]? & MASH-II &  \\
 327.1-02.2 & He 2-142 & 15 59 57.6 & -55 55 32.9 & [WC9] & AN2003 &  \\
 327.5-02.2 & MPA 1602-5543 & 16 02 11.2 & -55 43 30.0 & [WR] & MASH-II &  \\
 327.8+10.0 & NGC 5882 & 15 16 49.9 & -45 38 58.5 & Of(H) & SECGPN &  \\
 328.8-01.1 & PM 15 & 16 03 41.4 & -54 02 04.0 & [WR]? & MASH-I &   \\
 328.9-02.4 & He 2-146 & 16 10 41.2 & -54 57 32.9 & LST & JE1969 & bc-CSPN  \\
 329.0+01.9  & Sp 1 & 15 51 41.0 & -51 31 28.6 & O(H) & SECGPN & BL1990 \\
 329.5-02.2 & WRAY 17-75 & 16 12 34.4 & -54 23 35.3 & OB & t.w. &  \\
 329.8-02.1 & BMP 1613-5406 & 16 13 02.0 & -54 06 32.0 & Blue? & MASH-II & \\
 329.9+03.7 & BMP 1548-4936 & 15 48 52.1 & -49 36 48.0 & wels & MASH-II &  \\
 330.2+12.3 & BMP 1521-4228 & 15 21 26.0 & -42 28 08.0 & Blue & MASH-II & \\
 331.3+16.8 & NGC 5873 & 15 12 51.1 & -38 07 33.7 & wels & TA1993 &  \\
 331.4-03.5 & He 2-162 & 16 27 50.9 & -54 01 28.4 & O(H) & SECGPN &  \\
 331.7-01.0  & Mz 3 & 16 17 13.4 & -51 59 10.6 & B0 & LZ1998 &  \\
 331.8-02.3 & MPA 1624-5250 & 16 24 02.9 & -52 50  06.0 & wels? & MASH-II &  \\
 332.4-01.4 & BMP 1622-5144 & 16 22 34.0 & -51 44 56.0 & Blue & MASH-II & \\
 332.5-16.9 & HaTr 7 & 17 54 09.6 & -60 49 58.0 & DAO & SW1997 &  \\
 332.9-09.9 & He 3-1333 & 17 09 00.9 & -56 54 48.0 & [WC10] & AN2003 &  \\
 333.4-04.3 & PMR 3 & 16 41 04.4 & -53 02 24.0 & wels & AN2003 &  \\
 333.9+00.6 & PMR 5 & 16 19 40.2 & -49 13 59.0 & wels & AN2003 &  \\
 334.3-09.3 & IC 4642 & 17 11 45.0 & -55 24  01.5 & absorption lines & SECGPN &  \\
 334.8-07.4 & He 3-1312 & 17 03 02.9 & -53 55 54.0 & F(6-7) I; \newline  B[e]? & P2004; \newline  LZ1998 & bc-CSPN \\
 335.5+12.4 & DS 2 & 15 43 05.0 & -39 18 14.6 & O(H)  & SECGPN &  \\
 336.2-06.9 & PC 14 & 17 06 14.8 & -52 30  00.5 & [WO4] & AN2003 &  \\
 336.3-05.6 & He 2-186 & 16 59 36.1 & -51 42  06.5 & cont. & LS2007 &  \\
 336.5+05.5 & MPA 1611-4356 & 16 11 12.9 & -43 56 22.0 & [WO3] & MASH-II &  \\
 337.4+01.6 & Pe 1-7 & 16 30 25.9 & -46 02 50.8 & [WC9] & AN2003 &  \\
 337.5-05.1 & He 2-187 & 17 01 37.4 & -50 22 56.6 & OB & t.w. &  \\
 339.9+88.4 & LoTr 5 & 12 55 33.7 & +25 53 30.6 & G5 III & d2006 & FK1983 \\
 341.5+12.1 & Sand 3 & 16 06 28.4 & -35 45 13.0 & [WC3] & MN1982 &  \\
 341.6+13.7  & NGC 6026 & 16 01 21.1 & -34 32 36.6 & OB; \newline  WD/sdO & t.w.; \newline  d2009 & HB2006 \\
 341.8+05.4 & NGC 6153 & 16 31 30.8 & -40 15 14.2 & wels & LS2000 &  \\
 342.5-14.3 & Sp 3 & 18 07 15.8 & -51 01 10.1 & O3 & GP2001 & CB1999 \\
 343.5-07.8 & PC 17 & 17 35 41.1 & -46 59 51.3 & emission-line & t.w. &  \\
 343.6+03.7a & MPA 1644-4002 & 16 44 20.4 & -40 02 13.0 & Blue & MASH-II & \\
 343.9-05.8 & SB 30 & 17 27 02.3 & -45 32 38.5 & wels & GS2004 &  \\
 344.6-04.5 & MPA 1723-4419 & 17 23 06.1 & -44 19 16.0 & Blue & MASH-II & \\
 344.9+03.0 & BMP 1651-3930 & 16 51 41.3 & -39 30 27.0 & Blue & MASH-II & \\
 345.0-04.9 & Cn 1-3 & 17 26 11.8 & -44 11 29.1 & emission-line & t.w. &  \\
 345.2-08.8 & Tc 1 & 17 45 35.3 & -46 05 23.7 & Of(H) & SECGPN &  \\
 345.4+00.1 & IC 4637 & 17 05 10.5 & -40 53  08.4 & O(H) & SECGPN & CB1999 \\
 345.5+15.1 & Lo 13 & 16 09 45.9 & -30 55  07.6 & O(H) & SECGPN &  \\
 346.2-08.2 & IC 4663 & 17 45 28.5 & -44 54 11.5 & emission-line? & t.w. &  \\
 347.4+05.8 & H 1-2 & 16 48 54.1 & -35 47  09.1 & wels & GS2004 &  \\
 348.0-13.8 & IC 4699 & 18 18 31.2 & -45 59  03.2 & emission-line & t.w. &  \\
 348.4+04.9 & MPA 1655-3535 & 16 55 22.0 & -35 35 24.0 & [WC] & MASH-II &  \\
 349.3-01.1  & NGC 6337 & 17 22 15.7 & -38 29  03.5 & emission-line; \newline  M4 V & t.w.; \newline  d2006 & HB2006 \\
 349.5+01.0 & NGC 6302 & 17 13 44.2 & -37 06 15.9 & G V & d2006 & bc-CSPN \\
 349.7+04.0 & PPA 1702-3509 & 17 02 46.1 & -35 09 02.0 & [WR]? & MASH-I &   \\
 350.1-03.9 & H 1-26 & 17 36 29.7 & -39 21 57.0 & [WC4-5] & GS2003 &  \\
 350.9+04.4 & H 2-1 & 17 04 36.3 & -33 59 18.8 & [WC11]; \newline  O(H) & GC2009; \newline SECGPN &  \\
 351.1+04.8 & M 1-19 & 17 03 46.8 & -33 29 43.8 & wels? & AN2003 &  \\
 351.5-06.5 & SB 34 & 17 52 09.4 & -39 32 14.5 & [WO2] & GS2004 &  \\
 351.7-06.6  & SB 35 & 17 53 02.9 & -39 24  08.9 & wels & GS2004 &  \\
 352.1+05.1 & M 2-8 & 17 05 30.7 & -32 32  08.1 & [WO3] & GS2004 &  \\
 352.9+11.4 & K 2-16 & 16 44 49.1 & -28 04  04.7 & [WC11] & AN2003 &  \\
 352.9-07.5 & Fg 3 & 18 00 11.9 & -38 49 51.7 & cont. & t.w. &  \\
 354.7-03.9 & MPA 1748-3530 & 17 48 48.6 & -35 30 30.0 & Blue? & MASH-II & \\
 355.2-02.5 & H 1-29 & 17 44 13.8 & -34 17 33.1 & [WC4] & AN2003 &  \\
 355.3+03.8 & MPA 1719-3043 & 17 19 20.1 & -30 43 40.0 & Blue & MASH-II & \\
 355.4-04.0 & Hf 2-1 & 17 51 12.1 & -34 55 24.3 & [WO] & MA2009 &  \\
 355.7-03.5 & H 1-35 & 17 49 13.9 & -34 22 53.3 & emission-line? & t.w. &  \\
 355.9+03.6 & H 1-9 & 17 21 31.9 & -30 20 48.9 & [WC11] & GC2009 &  \\
 355.9-04.2 & M 1-30 & 17 52 58.9 & -34 38 23.0 & wels & AN2003 &  \\
 355.9-04.4 & K 6-32 & 17 53 40.3 & -34 43 41.0 & [WR]? & MASH-I &  \\
 356.0-04.2 & PHR 1753-3428 & 17 53 04.9 & -34 28 39.0 & [WR] & MASH-I &  \\
 356.1+02.7 & Th 3-13 & 17 25 19.4 & -30 40 42.0 & wels & GS2004 &  \\
 356.2-04.4 & Cn 2-1 & 17 54 33.0 & -34 22 21.2 & Of & AK1987 &  \\
 356.2-04.4 & Cn 2-2 & 17 54 33.0 & -34 22 21.2 & wels & TA1993 &  \\
 356.5-02.3 & M 1-27 & 17 46 45.5 & -33 08 35.1 & [WC11]? & GZ2006 &  \\
 356.5-03.9 & H 1-39 & 17 53 21.0 & -33 55 58.5 & [WC11]?  & GS2004 &  \\
 356.7-04.8 & H 1-41 & 17 57 19.1 & -34 09 49.1 & wels & TA1993 &  \\
 356.9+04.5 & M 2-11 & 17 20 33.3 & -29 00 39.1 & wels? & GC2009 &  \\
 357.1+03.6 & M 3-7 & 17 24 34.4 & -29 24 19.5 & wels & TA1993 &  \\
 357.1-04.7 & H 1-43 & 17 58 14.4 & -33 47 37.5 & [WC11] & GS2004 &  \\
 357.2-04.5 & H 1-42 & 17 57 25.2 & -33 35 42.9 & wels & GC2009 &  \\
 357.3+03.3 & M 3-41 & 17 25 59.8 & -29 21 50.4 & [WC11] & GC2009 &  \\
 357.6+01.7 & H 1-23 & 17 32 46.9 & -30 00 15.1 & Of? & AK1987 &  \\
 357.7-04.8 & BMP 1759-3321 & 17 59 45.2 & -33 21 13.0 & Blue & MASH-II & \\
 358.3-21.6 & IC 1297 & 19 17 23.5 & -39 36 46.4 & [WO3] & AN2003 &  \\
 358.8-00.0 & Te 2022 & 17 42 42.4 & -29 51 35.4 & OB & t.w. &  \\
 358.9-00.7  & M 1-26 & 17 45 57.7 & -30 12  00.6 & B[e]?; \newline  Of(H) & LZ1998; \newline SECGPN &  \\
 359.2+01.2 & 19w32 & 17 39 03.0 & -28 56 37.0 & symbiotic star? & BM2000 &  \\
 359.2-33.5 & CRBB 1 & 20 19 28.7 & -41 31 27.6 & O(H) & SECGPN &  \\
 359.3-01.8 & M 3-44 & 17 51 18.9 & -30 23 53.0 & [WC11] & GS2004 &  \\
 359.3-03.1 & M 3-17 & 17 56 25.6 & -31 04 16.8 & [WC11]?  & GS2004 &  \\
 359.4-05.6 & BMP 1807-3215 & 18 07 07.0 & -32 15 22.0 & Blue? & MASH-II & \\
 359.7+06.0 & BMP 1721-2554 & 17 21 58.1 & -25 54 24.0 & Blue & MASH-II & \\
 359.8+05.6 & M 2-12 & 17 24 01.5 & -25 59 23.3 & O; \newline  [WC11] & HP2007; \newline  GS2004 &  \\
\end{longtable}
\end{landscape}
}

\begin{landscape}
\begin{longtable}{p{2cm}p{3.3cm}p{2.5cm}p{2.5cm}p{3cm}p{2cm}p{2cm}}
\caption{\label{sample1} 
Catalogue of CSPN (possible PN)}\\
\hline\hline
  PN G  &  name  &  RA  &  DEC  &  classif. &  ref.  &  binary ref.   \\
\hline
\endfirsthead
\hline\hline
  PN G  &  name  &  RA  &  DEC  &  classif. &  ref.  &  binary ref.   \\
\hline
\endhead
\hline
\endfoot
 - & KPD 0005+5106 & 00 08 18.4 & +51 23 19.0  & DO & D1999 &  \\
 - & PG 0108+101   & 01 11 06.6 & +10 21 39.2  & DO & D1999 &  \\
 - & PG 0109+111   & 01 12 22.9 & +11 23 36.6  & DO & D1999 &  \\
 - & RWT 152   & 07 29 58.4 & -02 06 37.5 & sdO; \newline  O4-5 V & KC1989; \newline  C1980 &  \\
 - & PG 1034+001   & 10 37 04.0 & -00 08 20.0  & DO & HI2004 &  \\
 - & PG 1520+525   & 15 21 46.6 & +52 22  04.0 & PG 1159 & WH2006 &  \\
 - & He 2-139      & 15 54 44.5 & -55 29 34.1  & B[e] & LZ1998 &  \\
 - & PM 1-322      & 20 14 50.9 & +12 03 50.0  & symbiotic star? & MV2010 & \\      
 - & BD+28 4211    & 21 51 11.0 & +28 51 50.4  & O(H) & N1999 &  \\
\end{longtable}
\end{landscape}


References to Table~\ref{ample2} and \ref{sample1}:

\footnotesize{

\begin{description}
\item A1966 -  Abell 1966
\item AK1985  -  Aller  \& Keyes     1985
\item AK1987  -  Aller  \& Keyes     1987
\item AN2003  -  Acker \& Neiner      2003
\item CB2008 - Bilikova et al. 2008
\item BC1999 - Bond \& Ciardullo 1999
\item BC2003 - Benetti et al. 2003
\item B2008   -  Bohigas     2008
\item BD1993  -  Bianchi \& Defrancesco      1993
\item BG1987 -  Bond  \& Grauer 1987
\item BL1990 - Bond \& Livio 1990
\item BM1993 - Bond et al. 1993
\item BM2000  -  Belczyski  et al.      2000
\item BO2002 -  Bond et al. 2002
\item BP2002  -  Bond \& Pollacco      2002
\item BP2003  -  Bond et al.     2003
\item CB1999 - Ciardullo et al. 1999
\item C1980   -  Chromey     1980
\item C1995   -  Corradi      1995
\item CB1999  -  Ciardullo  et al.     1999
\item CG2009  -  Chu et al.     2009
\item CJ1987  -  Cohen \& Jones      1987
\item CP1985  -  Cerruti-Sola \& Perinotto     1985
\item D1983   -  Drilling    1983
\item D1985   -  Drilling 1985
\item D1999   -  Dreizler     1999
\item d2006   -  de Marco     2006
\item d2009   -  de Marco     2009
\item DB1996  -  Duerbeck \& Benetti      1996
\item dS2003  -  de Marco et al. 2003
\item EP2005 - Exter et al. 2005
\item F1999 -  Feibelman 1999
\item F1994   -  Feibelman    1994
\item FK1983  -  Feibelman \& Kaler     1983
\item FM1981 -  Ferguson et al. 1981
\item FP2006  -  Frew, Parker \& Russeil      2006
\item FS2010  - Frew et al. 2010
\item GB1983 -  Grauer \& Bond 1983
\item GB1987 -  Grauer et al. 1987
\item GC2009  -  G\'orny et al.     2009
\item GP2001  -  Gauba et al.    2001
\item GS2003  -  G\'orny \& Siódmiak      2003
\item GS2004  -  G\'orny et al.      2004
\item GT2000  -  G\'orny \& Tylenda     2000
\item GZ2003  -  Gesicki \& Zijlstra     2003
\item GZ2006  -  Gesicki et al.     2006
\item HB2006 -  Hillwig 2006
\item HI2006 -  Hsia et al. 2006
\item HZ2010  -  Hajduk et al.  2010
\item H2003   -  Handler     2003
\item HD1984  -  Heber \& Drilling     1984
\item HI2004  -  Hewett \& Irwin      2004
\item HP1993  -  Harrington \& Paltoglou      1993
\item HP2007  -  Hultzsch et al.     2007
\item JE1969  -  Jones, Evans \& Catchpole     1969
\item K1994   -  Kondrat'eva    1994
\item KB1994  -  Kingsburgh \& Barlow 1994
\item KB2005  -  Kraus et al.     2005
\item KC1989  -  Kwitter et al.     1989
\item L1977   -  Lutz  1977
\item LK1987  -  Lutz \& Kaler 1987
\item LS2000  -  Liu et al.     2000
\item LS2007  -  Lee et al.     2007
\item LR1983  -  Law \& Ritter
\item LZ1998  -  Lamers et al.     1998
\item MA2003  -  Marcolino \& de Araújo      2003
\item MA2009  -  Miszalski et al.     2009
\item MA2010  -  Miszalski 2010
\item MASH-I  -  Parker et al.  2006 
\item MASH-II -  Miszalski et al.     2008
\item MC2006  -  Mampaso et al.    2006
\item MC2010  -  Miszalski et al. 2010
\item MD1981  -  Margon et al. 1981
\item MK1988  -  M\'endez et al.    1988
\item MN1977  -  M\'endez \& Niemela 1977
\item MN1982  -  M\'endez \& Niemela      1982
\item MN1981  -  M\'endez \& Niemela 1981
\item MO2007  -  Mitchell et al.     2007
\item MP2001  -  Morgan, Parker \& Russeil     2001
\item MP2007  -  Mitchell et al.     2007
\item MV1997  -  Miranda et al.     1997
\item MV2010  -  Miranda et al.     2010
\item N1999   -  Napiwotzki     1999
\item NS1995  -  Napiwotzki \& Schoenberner      1995
\item NT2005  -  Napiwotzki et al. 2005
\item P1983   -  Pottasch    1983
\item P1996   -  Pottasch      1996
\item P2004   -  Pereira      2004
\item PF2004  -  Pierce et al.     2004
\item PM2002  -  Peña \& Medina     2002
\item PM2003  -  Parker \& Morgan      2003
\item PM2008  -  Pereira et al.  2008
\item PR1997  -  Pena et al.     1997
\item PT1992  -  Pena, Torres-Peimbert \& Ruiz      1992
\item RC2001  -  Rodríguez, Corradi \& Mampaso      2001
\item RD1998  -  Rauch, Dreizler \& Wolff      1998
\item RH2002  -  Rauch, Heber \& Werner      2002
\item RK1999  -  Rauch et al.      1999
\item SB2007  -  Smith,  Bally \& Walawender      2007
\item SECGPN  -  Acker et al. 1992
\item SF1987  -  Sabbadin, Falomo \& Ortolani      1987
\item SG2010  -  Santander-Garc\'ia 2010
\item SK1994  -  Stanghellini, Kaler \& Shaw     1994
\item SL2004  -  Shen, Liu \& Danziger     2004
\item SW1997  -  Saurer, Werner \& Weinberger      1997
\item SZ1997  -  Soker \& Zucker      1997
\item TA1993  -  Tylenda, Acker \& Stenholm      1993
\item TK1996  -  Tweedy \& Kwitter      1996
\item TN2004  -  Tovmassian et al. 2004
\item TS1987  -  Tamura \& Shaw      1987
\item WG2008  -  Weidmann et al.     2008
\item WH2006  -  Werner \& Herwig      2006
\item WK1997  -  Weinberger, Kerber \& Groebner     1997
\item WW1996  -  Walsh \& Walton 1996
\item WO1994  -  Wlodarczyk \& Olszewski 1994
\item ZP1990  -  Zijlstra, Pottasch \& Bignell     1990
\end{description}
}